\documentclass[%
 reprint,
superscriptaddress,
 amsmath,amssymb,
 aps,
]{revtex4-2}

\usepackage{graphicx}
\usepackage{dcolumn}
\usepackage{bm}

\usepackage{algorithm}
\usepackage{algorithmicx}  
\usepackage{algcompatible}
\usepackage{hyperref}
\usepackage{wrapfig}
\usepackage{amsmath,amssymb}
\usepackage{booktabs}
\usepackage{amsfonts}
\usepackage{amsthm}
\usepackage{comment}
\usepackage{color,soul}
\usepackage{xcolor,colortbl}
\usepackage{wrapfig}
\usepackage{float}
\usepackage{tabularx}
\usepackage{booktabs}
\usepackage{multirow}
\usepackage{xr}

\definecolor{Gray}{gray}{0.85}
\definecolor{LightCyan}{rgb}{0.88,1,1}
\newcolumntype{a}{>{\columncolor{Gray}}c}
\newcolumntype{b}{>{\columncolor{white}}c}

\DeclareMathOperator{\E}{\mathbb{E}}

\DeclareMathOperator{\Cov}{Cov}

\DeclareMathOperator{\argmin}{argmin}

\newcommand{\edit}[1]{{\color{black}#1}}
\newcommand{\newedit}[1]{{\color{black}#1}}
\newcommand{\revise}[1]{{\color{black}#1}}

\begin{document}

\preprint{APS/123-QED}

\newcommand{\change}[1]{\textcolor{blue}{#1}}

\title{Ab initio  uncertainty  quantification in   scattering analysis of microscopy}

\author{Mengyang Gu}
\email{Corres. authors: mengyang@pstat.ucsb.edu; yimin.luo@yale.edu}
 \affiliation{Department of Statistics and Applied Probability, University of California, Santa Barbara, Santa Barbara, CA 93106, USA}%

\author{Yue He}
\affiliation{Department of Statistics and Applied Probability, University of California, Santa Barbara, Santa Barbara, CA 93106, USA}%

\author{Xubo Liu}
\affiliation{Department of Statistics and Applied Probability, University of California, Santa Barbara, Santa Barbara, CA 93106, USA}%

\author{Yimin Luo}
\email{Corres. authors: mengyang@pstat.ucsb.edu; yimin.luo@yale.edu}
\affiliation{Department of Mechanical Engineering and Materials Science, Yale University, New Haven, CT 06511, USA}%

\date{\today}

\begin{abstract}
Estimating parameters from data is a fundamental problem in physics, customarily done by minimizing a loss function between a model and observed statistics. In scattering-based analysis, it is common to work in the reciprocal space.  Researchers often employ their domain expertise to select a specific range of wave vectors for analysis, a choice that can vary depending on the specific case. We introduce another paradigm that defines a probabilistic generative model from the beginning of data processing and propagates the uncertainty for parameter estimation, termed the \textit{ab initio} uncertainty quantification (AIUQ). As an illustrative example, we demonstrate this approach with differential dynamic microscopy (DDM) that  extracts dynamical information through minimizing a loss function for the squared differences of the Fourier-transformed intensities, at a selected range of wave vectors. We first show that \edit{the conventional way of estimation} in DDM is equivalent to fitting a temporal variogram in the reciprocal space using a latent factor model as the generative model. Then we derive the maximum marginal likelihood estimator, which optimally weighs the information at all wave vectors, therefore eliminating the need to select the range of wave vectors. Furthermore, we substantially reduce the computational cost of computing the likelihood function without approximation, by utilizing the generalized Schur algorithm for Toeplitz covariances. Simulated studies of a wide range of dynamical systems validate that the AIUQ method improves estimation accuracy and enables model selection with automated analysis. The utility of AIUQ is also demonstrated by three distinct sets of experiments: first in an isotropic Newtonian fluid, pushing limits of optically dense systems compared to multiple particle tracking; next in a system undergoing a sol-gel transition, automating the determination of gelling points and critical exponent; and lastly, in discerning anisotropic diffusive behavior of colloids in a liquid crystal. \edit{These studies demonstrate that the new approach does not require manually selecting the wave-vector range and enables automated analysis}. 

\end{abstract}
\maketitle
\setstcolor{blue}

\section{Introduction}
\label{sec:intro}

 Physical experiments play a crucial role in driving the advancement of basic science and technology. However, in recent years, the associated expenses and the laborious nature of data processing and analysis have also increased dramatically, posing obstacles to progress \cite{shang2019democratizing}. 
  One of the core challenges in this context is parameter estimation from data, conventionally performed by minimizing a loss function that quantifies the difference between the modeled and observed statistics. Notably, the estimation can depend critically on the choice of both the statistics and loss function applied to the fit, especially for spatiotemporally correlated measurements. 
  In the scattering analysis of dynamics, for instance, the range of wave vectors in the reciprocal space sometimes needs to be chosen
  in a case-by-case manner \cite{bayles2016dark,guidolin2023protein,martinez2012differential,lu2012characterizing,safari2017differential,richards2021particle,gao2015microdynamics}, which prohibits its use in high-throughput experiments. 
  
  In this work, we first introduce a new paradigm that defines a probabilistic generative model from the beginning of data processing. Then we propagate the uncertainty throughout the analysis by integrating out the \edit{random quantities}, to derive an optimal statistical estimator, such as the maximum marginal likelihood estimator, and subsequently implement it in fast algorithms. 
  Through the scattering analysis, we illustrate that the new estimator automatically weighs information from data at different bases, such as different wave vectors in Fourier-based analysis, thus lifting the barriers for selecting the wave-vector range.


We apply our automated and scalable analysis to video microscopy, one of the most ubiquitous tools to access the microscopic realm
, including cells, bacteria, and colloids.
Contemporary video microscopy offers not only visual insight but also versatility and power, supporting multiplexing imaging and capturing time sequences of dynamical processes. 
 Approaches to process video microscopy can be broadly defined in two classes:  particle tracking and basis decomposition tools.    
Particle tracking tools start with separating particle intensity profiles from the background for various techniques including fluorescence microscopy \cite{furst2017microrheology}, total internal reflection
fluorescence microscopy \cite{ewers2005single}, and dark field microscopy \cite{ueno2010simple,meng2021micromirror}. Once the particle profiles have been identified in individual frames, the trajectories of particles between consecutive time frames are linked using multiple particle tracking (MPT) algorithms, such as the Crocker-Grier Algorithm \cite{crocker1996methods}, 
also available in MATLAB \cite{blair2008matlab} and Python \cite{allan2016trackpy}. 
ImageJ \cite{schneider2012nih} and  Trackmate \cite{tinevez2017trackmate}, have been extensively used for analyzing biophysical and cellular processes. Machine learning tools have also been developed for static cellular images, such as the cell profiling tools (\textit{e.g.} CellProfiler \cite{carpenter2006cellprofiler}) and segmentation tools (\textit{e.g.} CellPose \cite{stringer2021cellpose}).
While there are well-established frameworks that explain static and dynamic errors, such as those proposed by \cite{savin2005static,savin2005role},  MPT algorithms still depend on several user-specified parameters for localization and selection, \textit{e.g.} the search radius for linking particle trajectories. As a result, particle-based tools can yield unsatisfactory results owing to factors such as irregular shapes, optically dense environments, fluctuations in the size, shape, and fluorescent intensity of moving objects.
On the other hand, basis decomposition tools reconstruct microscopy images through basis functions. \newedit{A class of basis decomposition methods termed Digital Fourier Microscopy (DFM) performs temporal analysis of the Fourier-transformed microscopy image sequences to extract process dynamics \cite{giavazzi2014digital}. This approach originates from scattering, which collects signals in the far field and computes the ensemble averages over many scatterers. It has been shown that near-field microscopy images encode the information on the correlation properties of dynamical processes analogous to scattering \cite{cerbino2008differential,giavazzi2009scattering}.} 
Among them, a representative method is differential dynamic microscopy, which was introduced in \cite{cerbino2008differential,giavazzi2009scattering}. These seminal works innovatively treat each pixel \newedit{in the Fourier space transformed} from a microscopy video as probes in dynamic light scattering \cite{berne2000dynamic}, as such, they can extract dynamical information of the system by correlating photon counts at distinct time points. In DDM, the squared difference of Fourier-transformed intensity at any \newedit{pairs of} frames, \edit{often referred to as the image structure function}, is related to the intermediate scattering function \cite{furst2017microrheology}. 
DDM analysis of video microscopy does not require localizing the particles and linking their trajectory, \newedit{provides access to high-quality analysis of the dynamics and requires no specialized imaging source or setup}. Thus, it serves as a complement to tracking-based tools such as MPT. DDM has been applied to a broad range of soft materials and biological systems, including, for instance, bacteria motility \cite{wilson2011differential,martinez2012differential}, 
colloidal gels \cite{gao2015microdynamics}, viscoelastic processes \cite{bayles2017probe}, active filament dynamics \cite{lee2021myosin}, and protein gelation dynamics \cite{meleties2022high}.


However, obstacles remain, preventing DDM from achieving full automation in high-throughput settings. \edit{In DDM, one fits the image structure function, which often varies by several orders of magnitude across different wave vectors and furthermore, the data are correlated over different lag time for the same wave vector. As such, this requires specifying the wave-vector range or weighing information at wave vectors in a case-by-case manner. 
} 
The difficulty in selecting wave vector to analyze is common for other scattering approaches, such as dynamic light scattering \cite{berne2000dynamic,stetefeld2016dynamic}. Thus solving it inspires new approaches for a wide range of characterization techniques. \edit{Here, we develop a probabilistic latent factor model, which encodes the physics-informed intermediate scattering function as the covariance function of latent factors, and derive the maximum marginal likelihood estimator, therefore removing the need for selecting wave vectors and weighing information in fitting the image structure function. Utilizing fast computational algorithms, such as the generalized Schur algorithm, we also reduce the cost for computing the likelihood function without approximation. 
} 



Our contributions are threefold. First of all, we introduce a new probabilistic generative model of scattering analysis of microscopy in Sec. \ref{subsec:latent_factor}. 
We show that conventional estimation of the parameters in DDM is equivalent to fitting a temporal variogram in the Fourier space using the generative model in Sec. \ref{subsec:ddm_connection}. \edit{This connection enhances our understanding of the existing estimator in DDM by integrating intermediate scattering function into the temporal covariance of a latent factor model at each wave vector. We show the conventional practice by fitting the image structure function in DDM is equivalent to minimizing the average of the temporal variogram from our latent factor model. To achieve better efficiency of estimation and remove the need to select wave vectors}, we derive the maximum marginal likelihood estimator (MMLE) after integrating out the latent factor processes introduced in Sec. \ref{subsec:MMLE}. 
As information on each wave vector is weighed appropriately by the likelihood, one can utilize all wave vectors instead of tailoring the range for different systems \edit{for estimation}. Second, directly computing the MMLE is prohibitively slow due to computing inversion and log determinants of a large number of covariance matrices. 
By evoking the Toeplitz structure of the covariances in the Fourier space, we apply the generalized Schur algorithm \cite{ammar1987generalized} to accelerate the computation of the marginal likelihood function without approximation in Sec. \ref{subsec:generalized_Schur}, reducing the computational order from cubic to pseudo-linear (or linear with respective to a log multiplicative constant) scaling to the number of time points, without approximating the likelihood. We show in Sec. \ref{subsec:data_reduction} that computational cost can be further reduced with a principled way of data reduction. Taken together, for a typical microscopy video with $500\times 500$ pixels and $500$ time frames, our approach is more than $10^5$ times faster than the direct computation of the likelihood function. Third, the generative model and fast algorithm enable a wide range of applications, such as automated determination of gelation point, discussed in Sec. \ref{subsec:anisotropic}. Finally, the new probabilistic approach provides uncertainty quantification of the estimation, and the likelihood can be utilized to select physical models by data. We demonstrate the approach by simulation studies and three distinct types of experiments, including optically dense particles, high-throughput determinant of gelling point, and estimation of anisotropic processes, where the estimation is automated for these applications.


We refer to this approach, which provides a probabilistic generative model and propagates the uncertainty from the beginning of the data analysis, the \textit{ab initio}  uncertainty quantification approach. The phrase \textit{ab initio} used herein should not be confused with the first principles calculation in quantum physics \cite{parr1950molecular}, though the philosophy of the computation may share some commonalities.   Physical or machine learning approaches that minimize a loss function for parameter estimation can motivate the development of the corresponding generative model, and the generative model 
enables us to scrutinize the underlying model assumptions made in estimation, and to build a more efficient estimator for different techniques. 
\edit{We have developed publicly available software packages in both R \cite{AIUQ2024Rpackage} and MATLAB \cite{AIUQ2024MATLABpackage} for automated scattering analysis of microscopy videos, where users can either call a built-in model or supply their models for inverse estimation.}



\section{Background for scattering analysis of microscopy}
\label{sec:background}
We first introduce the background and analysis for Differential Dynamic Microscopy (DDM) \cite{cerbino2008differential,giavazzi2009scattering}, which converts real-space coordinates into wave vectors in reciprocal space, 
and computes image correlations akin to Dynamic Light Scattering (DLS), leading to its characterization as ``\edit{scattering analysis of microscopy}''. 
Notable advantages of DDM include 
its compatibility with particles of different shapes \cite{reufer2012differential}, fast-moving particles \cite{brizioli2022reciprocal}, the ability to track particles and fluctuations at high optical density 
below the diffraction limit \cite{bayles2016dark}.

To start, we consider a system of $M$ particles in a two-dimensional (2D)  space with $\mathbf x_m(t)=(x_{m,1}(t),x_{m,2}(t))^T$ being the 2D particle location of particle $m$ at time $t$, for $m=1,...,M$. The normalized Fourier-transformed intensity can be written as the sum of the particle positions in the reciprocal  space  
\begin{align}
\psi(\mathbf q,t)=\frac{1}{\sqrt{M}}\sum^{M}_{m=1}\exp(-i \mathbf q \cdot \mathbf x_m(t)),   
\label{equ:cont_FT}
\end{align}
where $i$ denotes the imaginary unit and $\mathbf q$ is a 2D Fourier basis set or a wave vector. Assume the particles do not interact with each other \cite{furst2017microrheology}. 
 The intermediate scattering function (ISF), an important function encapsulating the time evolution of particle self-correlation, is characterized by a vector of parameters $\bm \theta$, below, 
\begin{align}
f_{\bm \theta}(\mathbf q,\Delta t)&=\mbox{Cov}(\psi(\mathbf q,t), \psi^*(\mathbf q,t+\Delta t)) \nonumber\\
&=\Bigl< \frac{1}{M}\sum^M_{m=1} \exp\left( i\mathbf q \cdot \Delta \mathbf  x_m(t,\Delta t)\right) \Bigr>, 
\label{equ:ISF}
\end{align}     
where $\mbox{Cov}(\cdot,\cdot)$ denotes the covariance operator, $\psi^*$ is the complex conjugate of $\psi$, $ \Delta \mathbf x_m(t,\Delta t)=\mathbf x_m(t+\Delta t)-\mathbf x_m(t)$, and $\langle\cdot\rangle=\mathbb E[\cdot]$ is the ensemble or expectation over time $t$. Here we use the notation $\langle\cdot\rangle$ and $\mathbb E[\cdot]$  interchangeably, to make it understandable to both physics and statistics communities. The derivation of Eq.  (\ref{equ:ISF}) is given in Appendix A.

Various processes have a closed-form expression of ISF. A few examples of closed-form ISF are derived in Appendix A and summarized in Table \ref{tab:ISF_table}. For instance, for Brownian motion (BM) or diffusive processes, the intermediate scattering function is $f_{BM}(\mathbf q,\Delta t)=\exp(-q^2 \theta \Delta t)$, where $q=|| \mathbf q||$ and here the only parameter  $\theta$ of ISF is the diffusion coefficient. By the cumulant theorem \cite{koppel1972analysis}, the ISF can be approximately characterized by the mean squared displacement (MSD), discussed in Appendix A.  
\edit{Our new approach is generally applicable to all ISFs with a vector of parameters $\bm \theta$, and the ISFs are not necessarily approximated by the MSD. We intend to estimate both the parameters $\bm \theta$ and system properties such as  MSD, which can be related to viscosity, storage and loss modulus through the Generalized Stokes–Einstein Equation (GSER) \cite{mason2000estimating}.   
}

We denote the light intensity of pixel $\mathbf x$ at time $t$ to be $y(\mathbf x,t)$. In Eq. (\ref{equ:ISF}), the ISF is the ensemble average of the  2D  spatial Fourier representation of the displacements of the particles from time $t$ to  $t+\Delta t$. 
To relate the ISF to pixel intensity in the Cartesian space,  the  Fourier-transformed difference in image intensity is studied in DDM \cite{cerbino2008differential,giavazzi2009scattering}: $\Delta {\hat y}(\mathbf q, t, \Delta t)=\mathcal F(y(\mathbf x, t+\Delta t)- y(\mathbf x, t))$,  with $\mathcal F(\cdot)$ denoting the 2D discrete Fourier transformation computed by fast Fourier transformation (FFT) \cite{cooley1965algorithm}. The time ensemble of this  quantity, often referred to as the image structure function, $D(\mathbf q,\Delta t)=\langle\Delta {\hat y}(\mathbf q, t, \Delta t)^2\rangle$, 
is often  modeled below: 
\begin{equation}
D(\mathbf q,\Delta t)= A(\mathbf q)(1-f_{\bm \theta}(\mathbf q, \Delta t)) + \bar B, 
\label{equ:image_SF}
\end{equation}
where $A(\mathbf q)$ is the real-valued scalar of amplitude parameter for wave vector $\mathbf q$, $f_{\bm \theta}$ is the ISF defined in Eq. (\ref{equ:ISF}), $\bar B$ denotes with mean value of the noise term.  
For an isotropic process in a square field of view with $N$ pixels, 
we denote an index set $\mathcal S_j=\{(j'_1,j'_2): q^2_{j'_1,1}+q^2_{j'_2,2}=q^2_j\}$ for $j=1,...,J$, which contains the 
indices of the $j$th `ring' of the Fourier-transformed quantity with amplitude $q_j=\frac{2\pi j}{\Delta x_{min} \sqrt{N}}$, with $\Delta x_{min}$ being the pixel size, i.e. the length of a pixel in one coordinate, and $N$ is the number of pixels in one frame. In total, there are $J$ rings of Fourier-transformed intensities, leading to $J$ distinct ISFs for isotropic processes.  
Furthermore, assuming we have n time frames, denote $\mathbf D$ as the $J\times (n-1)$ matrix with $(j,k)$th term being the observed image structure function $D(q_j,\Delta t_k)=\langle\Delta {\hat y}(\mathbf q_{\mathbf j'}, t, \Delta t_k)^2\rangle$, where now the ensemble is over both time $t$ and indices within each ring $\mathbf j'=(j'_1,j'_2)\in \mathcal S_j$. Let $\mathbf D_{m}$ denote the model output of $\mathbf D$, where the $(j,k)$th entry of $\mathbf D_{m}$ is $ D_{m}( q_j,\Delta t_k)=A_{j}(1-f_{\bm \theta}( q_j, \Delta t_k)) + \bar B$ with $\bar B$ denoting the mean for the random quantity $B$.  In DDM,  the parameters are often estimated by minimizing a loss function between the observed and modeled image structure functions 
\begin{align}
&(\bm \theta_{est}, \mathbf A_{est}, \bar B_{est})=\underset{\bm \theta, \mathbf A_{1:J}, \bar B}{\argmin} \mbox{Loss}( \mathbf D_{m}, \mathbf D ), 
\label{equ:loss_Dqt}
\end{align}
with $\mathbf A_{est}=[A_{est,1},...,A_{est,J}]^T$ being a J-vector of amplitudes. A typical choice of the loss function is either the $L_2$ or $L_1$ loss. 
As the parameter space has a high dimension, one often fits the model separately for each wave vector from a selected range, \edit{distinct in each application} \cite{cerbino2008differential,martinez2012differential, kurzthaler2018probing,you2021two}, and then estimators at different wave vectors are averaged to obtain $\bm \theta$. 
Some variants of DDM \cite{giavazzi2018tracking,escobedo2018microliter,cerbino2017dark} use different pre-specified estimators for $\bar B_{est}$ and an unbiased estimator of $A_{est,j}=2\langle \hat y(\mathbf q_{\mathbf j'}, t)^2\rangle_{\mathbf j'\in \mathcal S_j,t}-\bar B_{est}$ to estimate $A_{j}$, leaving  $\bm \theta$ in the ISF the only parameters to be numerically optimized. 
A summary of the estimators $\bar B_{est}$ is introduced in \cite{gu2021uncertainty}. 
In \cite{bayles2017probe,gu2021uncertainty}, the ISF is approximated by MSD, \textit{i.e.} $f_{ \bm \theta}(q,\Delta t)\approx \exp\left(-q^2  \langle \Delta x^2(\Delta t) \rangle/4 \right)$, and  Eq. (\ref{equ:image_SF}) is directly inverted at each $\Delta t$ to obtain the estimator of MSD $\langle x^2(\Delta t)\rangle$. In \cite{edera2017differential},  the MSD is obtained through iterative optimization to reduce numerical instability. 

\edit{However, estimation by directly fitting or inverting the image structure function can depend on the range of wave vector selected. This is illustrated by simulating a simple diffusion process, shown in Figure \ref{fig:direct_inverstion_dqt} in Appendix B.}  
Almost all existing approaches in DDM fit the image structure function in Eq. (\ref{equ:image_SF}), relying on selecting a subset of wave vectors to analyze, whereas selecting and reweighing the information at different wave vectors can be 
hard for a new system.  This difficulty arises from the substantial variations in amplitudes $A_j$ and the correlation of the image structure function at different lag times. A principled way to properly aggregate information at different wave vectors can unlock the tremendous potential for DDM to attain complete automation in this process, yet this has not been fully realized so far. 

To solve the challenge of optimally weighing information at different wave vector, 
a key question must be answered: {What is the probabilistic model implicitly assumed for the real-space image intensity in DDM?}  We bridge the physical approach and a probabilistic model to answer this question in Section \ref{sec:latent_factor}.

\section{Latent factor processes of video microscopy}
\label{sec:latent_factor}
\subsection{A latent factor model of  isotropic processes}
\label{subsec:latent_factor}
Let us first direct our attention to isotropic processes, where the ISF is the same at each ring of pixels in the reciprocal space. Extension to anisotropic processes will be discussed in Section \ref{subsec:anisotropic}.  
Consider a latent factor model of an $N=N_1\times N_2$ pixels of  real-valued image intensity $\mathbf y(t)=[y(\mathbf x_1,t),...,y(\mathbf x_{N},t)]^T$ at time $t$  below
\begin{equation}
 \mathbf {y}(t)= \frac{1}{\sqrt{N} }\mathbf W^{*} \mathbf z(t) + \bm \epsilon(t), 
 \label{equ:SAM}
 \end{equation}
where  $\bm \epsilon(t)\sim \mathcal{MN}(\mathbf 0, \frac{\bar B}{2} \mathbf I_N)$ is an N-dimensional Gaussian white noise vector with variance $\frac{\bar B}{2}$ and $\mathbf I_N$ being the identity matrix  of N dimensions,  the  $N \times N$ matrix $\mathbf W^{*}$ is a 2D 
inverse Fourier basis (or complex conjugate of the Fourier basis), which relates the $N$ observations of an image at time $t$ from Cartesian space $\mathbf x=(x_1,x_2)^T$ to a set of random factor processes $\mathbf z(t)$ in the reciprocal space $\mathbf q=(q_1, q_2)^T$. The latent factor $\mathbf z(t)$ is an $N$ dimensional complex random vector: $\mathbf z(t)=\mathbf z_{re}(t)+i\mathbf z_{im}(t) $, with each random factor at $n$ time points independently following a zero-mean multivariate normal distribution: $\mathbf z_{\mathbf j',re}\sim \mathcal{MN}(\mathbf 0, \frac{A_j}{4}\mathbf R_j)$ and  $\mathbf z_{\mathbf j',im} \sim \mathcal{MN}(\mathbf 0, \frac{A_j}{4}\mathbf R_j)$ for $j=1,...,J$ and  
 for any index $\mathbf j'\in \mathcal S_j$. 
 The $(k_1,k_2)$th entry of $\mathbf R_j$ is characterized by ISF: $R_j(k_1,k_2)= f_{\bm \theta}(q_j, \Delta t_k)$ with $\Delta t_k=|k_2-k_1| \Delta t_{min}$ with $\Delta t_{min}$ being the interval between two consecutive time  frames. \revise{The correlation matrix $\mathbf R_j$ encodes the two-time correlation function in the Fourier space \cite{brown1997speckle}, with the $(k_1,k_2)$ term of the covariance being $\frac{A_j}{4}R_j(k_1,k_2) =\mbox{Cov}\left({\hat y}_{q_{\mathbf j'}}(t_{k_1}),  {\hat y}_{q_{\mathbf j'}}(t_{k_2})\right)$ and for a time-invariant process, the quantity reduces to the image correlation function with $\Delta t=|t_{k_2}-t_{k_1}|$  in \cite{giavazzi2009scattering}.} The {key} is that the correlation matrix $\mathbf R_j$ of the latent factor is formed by the ISF from the physical process. 
This means each entry of the real and imaginary random factors corresponds to one Fourier-transformed quantity, where the covariance is parameterized by the amplitude and intermediate scattering function: $\mathbb E[z_{\mathbf q,re}(t) z_{\mathbf q,re}(t+\Delta t) ]=\mathbb E[z_{\mathbf q,im}(t) z_{\mathbf q,im}(t+\Delta t)]=\frac{A(\mathbf q)}{4} f_{\bm \theta}(\mathbf q, \Delta t)$. Without loss of generality, we assume that $N_1=N_2=\sqrt{N}$, i.e. square image at each time frame.

\subsection{DDM is fitting the temporal variogram of the latent factor model in the reciprocal space}
\label{subsec:ddm_connection}

Here, we draw the connection between DDM in fitting the image structure function and the latent factor model in Eq. (\ref{equ:SAM}). 
Note that the normalized discrete Fourier basis $\mathbf W/\sqrt{N}$ is a unitary matrix, i.e.   $\mathbf W \mathbf W^{*}=N\mathbf I_{N}$. 
By multiplying $\mathbf W/\sqrt{N}$ on both sides of Eq. (\ref{equ:SAM}) and  splitting the transformed vector into the real and imaginary parts $\mathbf {\hat y}(t)=\frac{\mathbf W \mathbf y(t)}{\sqrt{N} }=\mathbf {\hat y}_{re}(t)+i\mathbf {\hat y}_{im}(t)$, we have
\begin{align}
 \mathbf {\hat y}_{re}(t)= \mathbf {z}_{re}(t) + \bm {\hat \epsilon}_{re}(t), \label{equ:y_re_hat}\\
  \mathbf {\hat y}_{im}(t)= \mathbf {z}_{im}(t) + \bm {\hat \epsilon}_{im}(t),
 \label{equ:y_im_hat}
 \end{align}
 where $\bm {\hat \epsilon}_{re}(t) \sim \mathcal{MN}(\mathbf 0, \frac{\bar B}{4} \mathbf I_{N})$ and $\bm {\hat \epsilon}_{im}(t) \sim \mathcal{MN}(\mathbf 0, \frac{\bar B}{4} \mathbf I_{N})$ are both multivariate normal distribution with a diagonal covariance $\frac{\bar B}{4} \mathbf I_{N}$; $\mathbf z_{re}(t)$ and $\mathbf z_{im}(t)$ are both $N$ dimensional random vectors, where each entry corresponds to one wave vector $\mathbf q$ in the reciprocal space at time $t$.

We denote ${\hat y}_{\mathbf q}(t)$ and ${\hat y}_{\mathbf q}(t+\Delta t)$ to be the Fourier-transformed quantities at time frame $t$ and $t+\Delta t$ respectively, both on the wave vector $\mathbf q$. Then, we decompose their difference into the real and imaginary parts: ${\hat y}_{\mathbf q}(t+\Delta t)-{\hat y}_{\mathbf q}(t)=\Delta {\hat y}_{re,\mathbf q}(t,\Delta t)+i\Delta {\hat y}_{im,\mathbf q}(t,\Delta t) $. 
Based on the sampling model in Eq. (\ref{equ:SAM}),  both $\Delta {\hat y}_{re,\mathbf q}(t,\Delta t)$ and $\Delta {\hat y}_{im,\mathbf q}(t,\Delta t)$ follow the same normal distribution:  $ \mathcal N\left(0, \frac{A(\mathbf q)}{2}\left(1-f_{\bm \theta}(\mathbf q,\Delta  t)\right) +\frac{\bar B}{2} \right)$, as derived in Appendix C. 
Based on this result, one can compute the expectation of the squared  difference  of the Fourier-transformed intensity  between two frames, at any wave vector $\mathbf q$ and time difference $\Delta t$: 
 \begin{align}
 &\E\left[ {({\hat y}_{\mathbf q}(t+\Delta t)-{\hat y}_{\mathbf q}(t))({\hat y}^{*}_{\mathbf q}(t+\Delta t)-{\hat y}^{*}_{\mathbf q}(t)) }  \right] \nonumber  \\ 
 =&A(\mathbf q) (1-f_{\bm \theta}(\mathbf q,\Delta t)) +\bar B,  
\label{equ:connection}
 \end{align}
 which is the mean of the image structure function in Eq. (\ref{equ:image_SF}), the statistics used for estimating parameters in  DDM. 
 Eq. (\ref{equ:connection}) means that if we assume the probabilistic model in Eq. (\ref{equ:SAM}), the expected value of the image structure function in Eq. (\ref{equ:image_SF}) is the expected value of  ${({\hat y}_{\mathbf q}(t+\Delta t)-{\hat y}_{\mathbf q}(t))({\hat y}^{*}_{\mathbf q}(t+\Delta t)-{\hat y}^{*}_{\mathbf q}(t)) }$, equivalent to a temporal variagram in the reciprocal space.  Fitting a spatial variogram in the real space was extensively studied by the statistics community  \cite{cressie1980robust,cressie1985fitting,cressie1993statistics}. 
 However, it can be strenuous to fit the temporal variogram at each wave vector $\mathbf q$ in the reciprocal space and optimally aggregate the estimators, since the variogram is correlated at each lag time and the amplitude parameter can be drastically different at distinct $\mathbf q$.  Thus, directly fitting the temporal variogram in the reciprocal space and aggregating the estimators could lead to unstable estimation.
Next, we will introduce the maximum marginal likelihood estimator of the parameters, which provides a natural and optimal way to aggregate information on each wave vector.

\subsection{Maximum marginal likelihood estimator}
\label{subsec:MMLE}
 We denote two $n$-vectors $\mathbf {\hat y}_{re,\mathbf j'}=[y_{re,\mathbf j'}(t_1),...,y_{re,\mathbf j'}(t_n)]^T$ and $\mathbf {\hat y}_{im,\mathbf j'}=[y_{im,\mathbf j'}(t_1),...,y_{im,\mathbf j'}(t_n)]^T$ to be the Fourier-transformed quantity at wave-vector index $\mathbf j'$ over all $n$ time frames. 
 Denote the total observations and latent factors by $\mathbf Y$ and $\mathbf Z$, respectively. We integrate out the random factors to obtain the marginal distribution of observations: $p(\mathbf Y\mid \bm \theta,\bm A_{1:J}, \bar B )=\int p(\mathbf Y\mid \mathbf Z, \bm \theta,\bm A_{1:J}, \bar B ) p(\mathbf Z\mid  \bm \theta,\bm A_{1:J}, \bar B) d \mathbf Z$.  The marginal likelihood of $J$ rings of Fourier-transformed quantity in the reciprocal space follows: 
  \begin{align}
 &\mathcal L\left(\bm \theta,\bm A_{1:J}, \bar B \right)\nonumber \\
 =&  \prod^{J}_{j=1} \prod_{\mathbf j' \in \mathcal S_j} p_{MN}\left(\mathbf {\hat y}_{re,\mathbf j'};\, \mathbf 0,\, \bm \Sigma_j \right) \times
 p_{MN}\left(\mathbf {\hat y}_{im,\mathbf j'};\, \mathbf 0,\, \bm \Sigma_j \right), 
 \label{equ:prod_density}
 \end{align} 
where $\bm \Sigma_j= \frac{A_j}{4} \mathbf R_{j}+ \frac{\bar B}{4} \mathbf I_n$. 
 Here $\mathcal S_j$ denotes the index set of the $j$th ring of isotropic processes, for $j=1,...,J$, and $p_{MN}(\mathbf s;\bm \mu, \bm \Sigma)$ denotes the density of an n-vector multivariate normal distribution at a  real-valued vector $\mathbf s$ with mean and covariance being $\bm \mu$ and  $\bm \Sigma$ respectively:  
 \begin{align*}
 p_{MN}(\mathbf s;\bm \mu, \bm \Sigma)&=(2\pi)^{-\frac{n}{2}}|\bm \Sigma|^{-\frac{1}{2}}  \times \\
 &\qquad \exp\left\{- \frac{1}{2}(\mathbf s-\bm \mu)^T \bm \Sigma^{-1}(\mathbf s-\bm \mu)\right\}.
 \end{align*}
 The derivation of Eq. (\ref{equ:prod_density}) is given in Appendix C. 

We denote $S_j=\#\mathcal S_j$, the number of transformed pixels within the index set $\mathcal S_j$, and let the total number of pixels within $J$ rings be $\tilde N=\sum^J_{j=1}S_j$. Note that we do not include the transformed output outside of the $J$ rings in the likelihood function, consistent with DDM, and hence $\tilde N<N$. In principle, one can compute the likelihood of all quantities inside or outside the $J$ rings. However,  as the transformed quantities at large wave vectors become similar to noise, we may only need to include a small number of wave vectors, which is much smaller than $J$. Based on this feature, we will introduce feasible ways to further reduce the computation and storage cost in Section \ref{subsec:data_reduction}. \edit{Using all wave vectors for estimation is implemented as the default version of our software packages \cite{AIUQ2024Rpackage,AIUQ2024MATLABpackage}; a reduction of the wave-vector range can be executed to decrease computational cost.}

Here one can maximize the marginal likelihood function in Eq.  (\ref{equ:prod_density}) to estimate parameters $\bm \theta$, $\mathbf A_{1:J}$ and $\bar B$. 
However, direct maximization can become unstable as $\mathbf A_{1:J}$ contains $J$ parameters which can be large. 
 The connection between the latent factor model in Eq. (\ref{equ:SAM}) enables us to understand the properties of some estimators of the parameters $A_{1:J}$ and $\bar B$. 
 For any $\bar B$, an unbiased estimator of the amplitude parameter  $A_j$ follows  
 \begin{equation}
  A_{est,j}=\frac{2}{ S_j n} \sum_{\mathbf j'\in \mathcal S_j}\sum^{n}_{k=1}|\hat y_{\mathbf j'}(t_k)|^2-{\bar B}
  \label{equ:A_est_j}
 \end{equation}
 for $j=1,...,J$. \revise{The unbiasedness of the estimator $A_{est,j}$ is derived in Appendix C}. In practice,  we may take the absolute value of $A_{est,j}$  to keep it nonnegative. \newedit{Another way is to set those negative $A_{est,j}$ to 0. As $A_{est,j}$ is rarely negative and the absolute value of the negative $A_{est,j}$ is typically very small, they almost have no impact on the likelihood function.  Thus, these two estimations do not change the parameter estimation significantly, as shown in Table S4 in the Supplemental Material.
 In our package, we offer users the option to use absolute values or setting $A_{est,j}$ to 0.}
Then we estimate the model parameter $\bm \theta$ and noise parameter $\bar B$ by maximizing the marginal  likelihood with the estimated $\mathbf A_{est, 1:J}$ from Eq. (\ref{equ:A_est_j})
\begin{equation}
(\bm \theta_{est},\bar B_{est})=\mbox{argmax}_{\bm \theta,\bar B} \mathcal \mbox{log} \left[L\left(\bm \theta,\mathbf A_{est,1:J},\bar B \right)\right].
\label{equ:log_lik_theta_B}
\end{equation}
The plug-in estimator $\mathbf A_{est,1:J}$  dramatically reduces the dimension of the parameter space and makes numerical optimization much more stable than optimizing in a large parameter space.   We use the low-storage quasi-Newton
optimization method \cite{nocedal1980updating} for estimation $\bm \theta$ and $\bar B$ by  maximizing the logarithm of the marginal likelihood function $\mbox{log}\left[\mathcal L\left(\bm \theta,\mathbf A_{est,1:J}, \bar B \right)\right]$ in Eq. (\ref{equ:log_lik_theta_B}).

Eq. (\ref{equ:SAM}) defines a generative model for the untransformed data from the beginning of the analysis, and the latent factor processes are integrated out to propagate the uncertainty to derive the  
marginal likelihood in Eq. (\ref{equ:prod_density}) for parameter estimation. 
Thus we call the method  \textit{ab initio uncertainty quantification} (AIUQ) for scattering analysis of microscopy.
Note here we do not need to compute the difference of image pairs as in the image structure function in Eq. (\ref{equ:image_SF}); instead, we will apply the generalized Schur algorithm \cite{gohberg1972inversion,ammar1988superfast, ling2019superfast} for accelerating the computation of Toeplitz covariances from the marginal likelihood in Eq. (\ref{equ:prod_density}) summarized in Section \ref{subsec:generalized_Schur}  and Section S1 in SI. \newedit{The AIUQ approach provides a global fit of the entire data at the untransformed real space and time $\{\mathbf x, t\}$ based on the  model in Eq. (\ref{equ:SAM}). In comparison, DDM fits the image structure function in the Fourier space in Eq.  (\ref{equ:image_SF}) at the lag time $\{\mathbf q, \Delta t\}$, and selecting suitable wave-vector ranges may be required to obtain an accurate estimation of the ISF.   }

There are two ingredients of the AIUQ approach for scattering analysis of microscopy. The first key part is to model the untransformed intensity by a probabilistic model, where the temporal correlation of each latent factor process is parameterized by an intermediate scattering function at each wave vector. Equivalently, this is to assume a Bayesian prior encoded physics information of the process. The second key ingredient is to integrate (or marginalize) out random factor processes and estimate the governing physical parameters by an asymptotically optimal estimator--the maximum marginal likelihood estimator.  
Integrating out random quantities in the model is the foundation of Bayesian analysis, which inherently avoids overfitting the data and provides uncertainty quantification. Readers are referred to \cite{berger1999integrated,lakshminarayanan2017simple,wilson2020bayesian,gu2022scalable} for more discussion and applications of the marginal likelihood in machine learning and Bayesian analysis.

 \begin{figure} 
\includegraphics[width =0.50\textwidth]{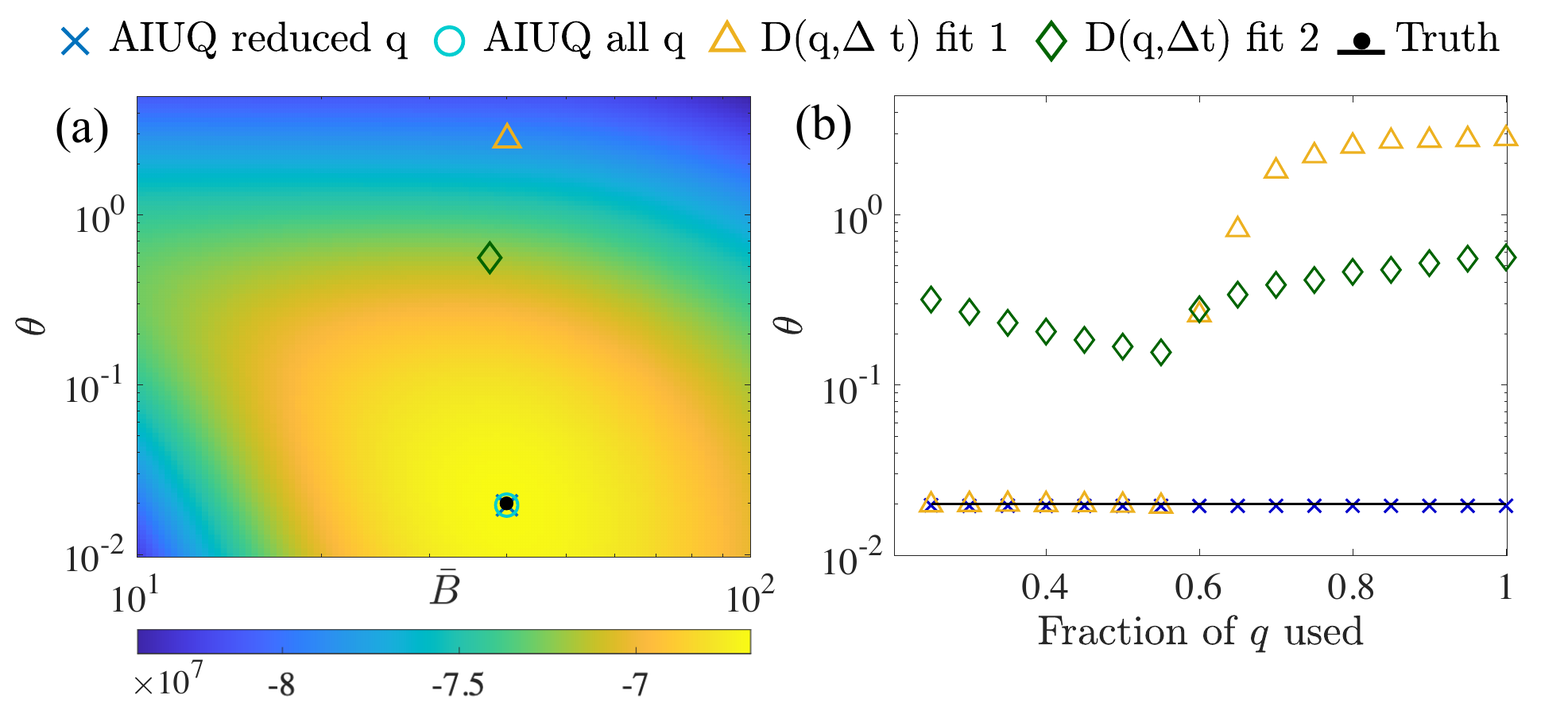}
   \caption{(a) \newedit{For a simulated image stack,} B, $\theta$ estimates 
   using different methods: AIUQ (x), AIUQ with all q (o), $D(q,\Delta t)$ fit 1 ($\triangle$), $D(q,\Delta t)$ fit 2 ($\diamond$), whereas the solid dot denotes the truth. The colormap represents the log marginal likelihood by the AIUQ approach with reduced $q$. (b) The  estimated diffusion coefficient $\theta$ using different methods with the fraction of $q$ used in estimation. The axes are plotted on logarithmic scales. }  
   \label{fig:MSD_pos_record_true_B_40_true_D_0_02}
\end{figure}

\subsection{Comparison between estimation by maximum marginal likelihood estimator and fitting image structure function}
\label{subsec:comparison_mmle_fitting}

We illustrate the estimation accuracy of AIUQ using a slow diffusive process in Fig. \ref{fig:MSD_pos_record_true_B_40_true_D_0_02}. When dealing with slow dynamics, where the standard deviation of the displacement is smaller than 1/10 of a pixel in each timestep,  MPT sometimes fails to accurately capture the underlying dynamics at small lag times, due to the similarity between noise and signal. 
We compare AIUQ with two DDM approaches both based on all wave vectors. In the first DDM approach, termed \edit{$D(q,\Delta t)$ fit 1}, the noise parameter is estimated by $\bar B_{est}=D(\mathbf q_{min},\Delta t_{min})$, and the amplitude parameter  $A_j$ is estimated by the unbiased estimator specified in Eq. (\ref{equ:A_est_j}). Then these parameters are substituted into Eq. (\ref{equ:loss_Dqt}), so that the parameters in ISF \revise{$\bm \theta$} are numerically optimized by minimizing the $L_2$ loss between the reconstructed and observed image structure functions \revise{by summing up the loss at all wave vectors}.   In the second DDM approach, termed \edit{$D(q,\Delta t)$ fit 2},  parameters are separately estimated for wave vectors by numerically minimizing the $L_2$ loss function of the image structure function in Eq. (\ref{equ:loss_Dqt}) for each $q_j$ \revise{to obtain $\bm \theta(q_j)$}, and then the average of the estimation of \revise{$\bm \theta_j(q_j)$} is used to estimate the parameters \revise{$\bm \theta$}.   \edit{In both approaches, we find that the estimation of noise parameter $\bar B$ is reasonably good (Fig 1 (a)). 
Furthermore, $\bar B$ is also reasonably good compared with the truth in a wide range of simulation studies shown in Table S3 in the Supplemental Material.
However, the estimation of diffusion coefficient by two DDM approaches is not satisfying when the entire wave-vector range is used in estimation.}
AIUQ approaches, with all wave vectors or with a reduced set of wave vectors, use the first $J_0$ sets that explain $99\%$ of the variability:  $({\sum^{J_0}_{j=1} A_{j}})/({\sum^J_{j=1} A_j})\geq 99\%$, are compared. 
As demonstrated in Fig.  \ref{fig:MSD_pos_record_true_B_40_true_D_0_02}(a), AIUQ is more accurate \edit{in estimating the  diffusion coefficient than two approaches that fit the image structure function, while performing similarly well in estimating the noise parameter.}  Fig. \ref{fig:MSD_pos_record_true_B_40_true_D_0_02}(b) shows fitting the image structure function in DDM can lead to q-dependent estimation. \edit{We further show another example when we directly invert the image structure function in Fig. \ref{fig:direct_inverstion_dqt} for estimation, which demonstrated the q-dependence for estimation}.
In comparison, the estimated parameter is not dependent on $q$ using AIUQ, when a sufficient number of wave vectors is used, shown in Fig. \ref{fig:MSD_pos_record_true_B_40_true_D_0_02}(b), as the information is naturally weighed 
by the marginal likelihood in Eq. (\ref{equ:prod_density}).  \newedit{Selecting a suitable wave-vector range for estimation can improve the estimation by DDM, whereas a principled way of selecting  such range was not found}.

\subsection{Extension to  anisotropic processes}
\label{subsec:anisotropic}

DDM has facilitated the analysis of anisotropic fluctuations in liquid crystals \cite{giavazzi2014viscoelasticity},  estimation of directional motions in flow \cite{richards2021particle} and in response to a magnetic field  \cite{pal2020anisotropic}. 
To address angle-specific or anisotropic motion, previous works select a limited angle range centered around the desired direction. For instance, in \cite{reufer2012differential}, a specific range of $q$ values is averaged to extract distinct translational Brownian diffusion along perpendicular axes of ellipsoids. Another example is liquid crystal, explored in \cite{giavazzi2014viscoelasticity}, where a bowtie-shaped $q$-space region along both parallel and perpendicular directions to the director is averaged to extract viscoelastic properties and isolate polarized light contribution. However, constraining $q$ in a certain angle range could discard useful information on other wave vectors outside this range.

The AIUQ approach of scattering analysis of microscopy can be extended to anisotropic processes. Let us consider a 2D anisotropic process, 
we can split the intermediate function to two coordinates: $f_{\bm \theta}(\mathbf q, \Delta t)=f_{\bm \theta_1}(q_1,\Delta t) f_{\bm \theta_2}(q_2,\Delta t)$, where $f_{\bm \theta_l}(q_l,\Delta t)$ is an intermediate scattering function for the $l$th coordinate, and the parameters can be split to $\bm \theta=\{\bm \theta_1, \bm \theta_2\}$ with $\bm \theta_1$ and $\bm \theta_2$ being the parameters of $f_{\bm \theta_l}(q_l,\Delta t)$  for $l=1,2$.  From the cumulant approximation in \edit{Eq. (\ref{equ:cumulant_ISF})} discussed in Appendix A, one derives the anisotropic ISF
\begin{equation*}
f_{\bm \theta}(\mathbf q, \Delta t)\approx\exp\left(-\frac{q^2_1 \langle \Delta x^2_1(\Delta t)\rangle+ q^2_2 \langle \Delta x^2_2(\Delta t)\rangle}{2}\right),
\end{equation*}
where 
$\langle \Delta x^2_1(\Delta t) \rangle$ and $\langle \Delta x^2_2(\Delta t) \rangle$ are MSD at $\Delta t$ along the two coordinates, respectively. As the process is anisotropic, the ISF is different for each wave vector in general. Then one can compute the maximum marginal likelihood estimator in Eq. (\ref{equ:log_lik_theta_B}) with the anisotropic ISF. Since the amplitude $A_j$ depends on the transformed intensity at zero lag time and image noise \cite{giavazzi2009scattering,nixon2022probing}.
Thus, $A_j$ can be calculated similarly regardless of isotropic or anisotropic processes.  The projected intensity on all wave vector can be used in an AIUQ approach, which makes the analysis more stable and accurate.

\section{Fast estimation, uncertainty assessment and model selection}
\label{seq:computation}
\subsection{The generalized Schur method of accelerating computation for Toeplitz covariances}
\label{subsec:generalized_Schur} 
Directly computing the marginal likelihood in Eq. (\ref{equ:prod_density}) is slow as computing each of $n$-vector multivariate normal density requires $\mathcal O(n^3)$ computational operations, from computing the matrix inversion and determinant of the covariance matrix.  Assuming we have $J$ rings of Fourier-transformed intensity, and the number of indices in each ring is $S_j$,  the total computational operations scales as   $\mathcal O(Jn^3)+ \mathcal O(\tilde N n^2)$ operations for isotropic processes, and $ \mathcal O(\tilde N n^3)$ for anisotropic processes. As the likelihood function needs to be computed tens of times for numerically optimizing the parameters, the computational operations can be up to $10^{15}$ for a regular video with  $500\times 500$ pixels and $500$ time frames, which is too computationally expensive.   
Here, we introduce a fast approach that can substantially reduce the computational cost without any approximation to the likelihood function.  

Note that as video microscopy is often taken equally spaced in time, which means the covariance matrix $\bm \Sigma_j$ is a Toeplitz matrix \cite{gray2006toeplitz} for each j, parameterized by the ISF: 

\begin{equation}
{\bm \Sigma}_j =
\begin{bmatrix}
\tilde f_{j,0} &  \tilde f_{j,1} & \tilde f_{j,2} &\dots &\tilde f_{j,n-1}\\
\tilde f_{j,1}  & \ddots & \tilde f_{j,1} &\dots &\tilde f_{j,n-2}\\
\tilde f_{j,2}  &\tilde f_{j,1} &\ddots &\ddots &\vdots\\
\vdots &\vdots &\ddots &\ddots &\tilde f_{j,1}  \\
\tilde f_{j,n-1} &\tilde f_{j,n-2} &\dots &\tilde f_{j,1} &\tilde f_{j,0}\\
\end{bmatrix},
\label{eq:Toeplitz}
\end{equation}
where $\tilde f_{j,k}=\frac{A_j}{4} f_{j,k}+\frac{\bar B}{4} 1_{k=0}$ with $f_{j,k}=f_{\bm \theta}(q_j,\Delta t_k)$ being the intermediate scattering function at ring $j$ and $\Delta t_k$, and $1_{k=0}$ being a Delta function at $k=0$. Consequently, the covariance matrix $\bm \Sigma_j$ is a Toeplitz matrix. 
The generalized Schur algorithm 
was developed in \cite{ammar1987generalized, ammar1988superfast} for accelerating the computation,  reducing the computational cost of inversion and log determinant of Toeplitz covariance from $\mathcal O(n^3)$ to $\mathcal O(n(\log(n))^2)$ operations for an $n\times n$ Toeplitz covariance matrix. Thus, to compute the marginal likelihood function in Eq. (\ref{equ:prod_density}), one only needs $\mathcal O(Jn(\log(n))^2)+\mathcal O(\tilde N n \log(n))$ operations for isotropic processes, where the first term is from computing determinant and matrix inversion of $J$ Toeplitz covariances, and the second term is from computing the $\tilde N$ Toeplitz matrix-vector multiplication, through FFT. For anisotropic processes, one only needs $\mathcal O(\tilde N n(\log(n))^2)$ operations for computing the likelihood.  
The generalized Schur method of computing inversion, log determinant, and matrix-vector multiplication for  Toeplitz matrices was implemented in the `SuperGauss' algorithm in R platform \cite{Ling202SuperGauss,ling2019superfast}. The generalized Schur algorithm is a `superfast' algorithm as its complexity is pseudo-linear to the number of observations for decomposing a Toeplitz covariance \cite{ammar1988superfast}, and other algorithms, such as the Levinson-Durbin algorithm \cite{levinson1946wiener} that take $\mathcal O(n^2)$ operations for decomposing a Toeplitz covariance, are generally called fast algorithms. \edit{Details of the generalized Schur algorithm are discussed in SI Section S1.}

\subsection{Data reduction}
\label{subsec:data_reduction} 
 As the marginal likelihood function in Eq. (\ref{equ:prod_density}) often needs to be computed tens of times to iteratively find the maximum value, it is of interest to further reduce the computational complexity for large videos in addition to using the generalized Schur algorithm. 

For processes with an increased MSD with respect to the increase of lag time, such as subdiffusion or superdiffusion,  the transformed intensity rapidly decorrelates at large wave vectors, making it indistinguishable from noise.  A typical way for dimension reduction is to choose the first $J_0$  rings of Fourier-transformed intensity that explains the most variability of the data: $({\sum^{J_0}_{j=1} A_{j}})/({\sum^J_{j=1} A_j})\geq 1-\varepsilon$, where $\varepsilon$ is a small number, similar to the probabilistic principal component analysis \cite{tipping1999probabilistic}. 
For instance, choosing $\varepsilon=0.001$ means we have $99.9\%$ of the variability from the transformed intensity explained. As $A_j$ becomes close to $0$ at large $j$ and the `ring' gets larger for large $j$,  such a choice  can substantially reduce the storage and computational requirements by avoiding computing a large number of transformed intensities at high-frequency wave vectors. Another way is to ensure that we select a relatively large proportion ($J_0/J\geq \beta$) to retain enough information from the signal. For some confined processes, such as the Ornstein–Uhlenbeck  (OU) process, where the MSD approaches a plateau at large lag time,  using a large fraction of wave vectors is a safer choice, as the temporal correlation of transformed intensities does not decrease at large wave vectors.   
Having a more conservative threshold that selects a larger $J_0$, e.g. $\varepsilon=0.001$ and $\beta=0.5$, is also better for estimating the noise parameter. Here, we emphasize that the selection of $J_0$ is mainly for further reducing the storage and computational cost, which is different from selecting the range of wave vectors to analyze in minimizing the loss function in Eq. (\ref{equ:loss_Dqt}) in DDM. The default version of our packages utilize all wave vectors in estimation \cite{AIUQ2024Rpackage,AIUQ2024MATLABpackage}.

\subsection{Confidence interval}
The uncertainty of parameter estimation can be quantified with the availability of the probabilistic generative model in Eq. (\ref{equ:SAM}).  
 Here, estimation errors stem from two sources: discretization of pixels and parameter estimation. First, the dynamical processes are continuous, whereas analysis is performed on discretized pixels. 
 The pixel-related uncertainty can be big when we have a small number of pixels, and the error becomes small when we have a finer pixel size.  Second, similar to all statistical inferences, the stochastic nature of observations introduces uncertainty in parameter estimation. Given a generative model, the uncertainty from statistical analysis can be assessed by either the Frequentist  or Bayesian analysis. When distributions of the pivotal quantities have no closed-form expressions, the confidence interval from Frequentist analysis is often approximated by the central limit theorem \cite{mardia1984maximum} or bootstrap \cite{davison1997bootstrap}. The posterior credible interval from Bayesian analysis is by specifying the prior of the parameters and then computing the posterior distribution, often evaluated by the Markov chain Monte Carlo algorithm \cite{hoff2009first}, such as Gibbs sampling and Metropolis algorithm \cite{geman1984stochastic,metropolis1953equation}.  
 The parameter uncertainty of Bayesian and Frequentist analysis typically agrees when the sample size approaches infinity. 

We quantify both the pixel and estimation uncertainty to construct the confidence interval with the scalable algorithm by the generalized Schur method. To integrate the pixel uncertainty, we compute the maximum marginal likelihood estimator of $(\bm \theta, \bar B)$ by letting the associated amplitude of the wave vector $q_j$ to be $q_j- \Delta  q_{min}$  and $q_j-\Delta q_{max}$ for $j=1,...,J$ separately. Then we follow \cite{mardia1984maximum} to approximate the parameter estimation uncertainty through an asymptotically normal distribution. Notably, the scattering information comes from $M$ trajectories ($M \approx$ 50-200) of individual particles instead of $\sum^{J_0}_{j=1} S_j$ Fourier-transformed time series, and typically  $M \ll N $ ($N \approx 2.5\times 10^5$). Thus, one needs to discount the likelihood by a power of $M/N$ when computing the uncertainty in asymptotic normality approximation. We integrate both sources of the uncertainty to construct the confidence interval of estimated parameters. 
As will be shown in the simulated studies, the $95\%$ confidence intervals are narrow when the size of the image is not too small, and they cover the truth most of the time.  

\subsection{Model selection and diagnostics}
\label{subsec:model_selection}



Given a few plausible models of ISF, how do we know which one shall be used? \edit{The conventional way in DDM is to evaluate the ``fit'' of the image structure function. However, as each transformed datum in the image structure function exhibits different variances and is correlated, it may be hard to select a metric to evaluate the fit. Instead, one may focus on the ``fit'' in the original Cartesian space.} Statistical information criteria, such as  Akaike information criterion (AIC)   \cite{akaike1998information}, may be computed to evaluate the fit in original space with the maximum likelihood estimator: $\mbox{AIC}=2p-\mbox{log}(\mathcal L(\bm \theta_{est},\mathbf A_{1:J,est},\bar B_{est}))$, where $\mathcal L(\bm \theta_{est},\mathbf A_{1:J,est},\bar B_{est})$ denotes the maximum likelihood value and $p$ is the number of parameters to be estimated. AIC quantifies the predictive error, and hence one selects a model with a small AIC. In practice,  models with a larger set of parameters are often selected by AIC, when the number of observations is large. One can compensate by using the Bayesian information criterion (BIC) \cite{schwarz1978estimating}, which also penalizes for the number of observations for model selection. 

Alternatively, we can compute the one-step predictive error to evaluate the fit for model selection. To do so, we first divide the microscopy video into two groups, and the first $n_0$ time frames are used for estimating parameters. Then we make predictions sequentially on each $n^*=n_0+1,...,n$ using the one-step-ahead prediction  
\begin{equation}
\mathbf y_{pred}(t_{n^*})=\mathbf W^{*} \mathbf z_{pred}(t_{n^*}), 
\label{equ:y_pred_t}
\end{equation}
\newedit{where for isotropic processes, the $\mathbf j'\in \mathcal S_j$ column of $\mathbf z_{pred}(t_{n^*})$ is 
\begin{align*}
z_{pred,\mathbf j'}(t_{n^*})&= \mathbf r^T(t_{n^*}) ( \mathbf R_{j} + \eta_{est}\mathbf I_{n^*-1})^{-1} \hat {\mathbf y}_{\mathbf j'}(\mathbf t_{1:(n^*-1)}), 
\end{align*}
with $\mathbf r(t_{n^*})=(f_{\bm \theta_{est}}(q_j, t_{n^*}-t_1),...,f_{\bm \theta_{est}}(q_j,t_{n^*}-t_{n^*-1}) )^T$ is an $n^*-1$ vector of ISF, and $\mathbf R_{j} $ is a $({n^*}-1)\times ({n^*}-1)$ matrix with the $(k,k')$th entry of being $f_{\bm \theta_{est}}(q_j,t_k-t_{k'})$, and  $\eta_{est}=\bar B_{est,j}A^{-1}_{est,j}$. }
The generalized Schur algorithm is used for accelerating the computation of predictive mean in Eq. (\ref{equ:y_pred_t}) to avoid the direct inversion of the covariance matrices.  
Then we select the model that minimizes the predictive loss, such as the average root mean squared error. 
When predictive errors are similar, a model with a smaller number of parameters is preferred.

  

 \begin{table*}[t]
\begin{tabular}{l|c|c|c|c|c|c|c|c}
 \toprule
Processes,&  \multicolumn{4}{c|}{Small video}
  &  \multicolumn{4}{c}{Regular video}
  \\
   \cmidrule{2-9}
true parameters  & $D(q,\Delta t)$  & $D(q,\Delta t)$  & AIUQ & AIUQ & $D(q,\Delta t)$  & $D(q,\Delta t)$   & AIUQ & AIUQ \\ 
 & fit 1 & fit 2 &  reduced q &all q & fit 1 & fit 2 & reduced q &all q \\
 \midrule

 BM, $\sigma^2_{BM}=.020$  &8.3 & .36&{\textbf{.020}} (.017,.024] &{\textbf{.020}} (.017,.024] &2.8 &.56 & .019 [.019,.020]  &{\textbf{.020}}  [.019,.020] \\ 
\midrule
  BM, $\sigma^2_{BM}=2.0$   &3.8 & 1.7&{\textbf{2.0}} [1.8, 2.4]& 2.1  [1.8, 2.4] &3.6 &2.8 & {\textbf{2.0}} [2.0, 2.1]   &{\textbf{2.0}}  [2.0, 2.1] \\
\midrule
FBM, $\sigma^2_{FBM}=8.0$  &7.5 & 4.2 &{\textbf{8.0}} (7,10] &{\textbf{8.0}} (7,10] &6.7 &3.8 & {\textbf{8.1}} [7.8,8.5)  &{\textbf{8.1}}  [7.8,8.5)\\
$\alpha=.60$ &.75 & 1.0&
{\textbf{.59}} [.35,.67) &{\textbf{.59}} [.35,.67)  &.87 &.88 & {\textbf{.59}} [.58,.61) &{\textbf{.59}} [.58,.61) \\
\midrule
FBM, $\sigma^2_{FBM}=.50$  &2.3 & \textbf{.49} & .52 (.42,.52]& .52 (.41,.52]& 3.0& .74&  \textbf{.50} [.47,.54] & \textbf{.50} [.46,.55] \\
$\alpha=1.4$ &1.3 & 1.3& \textbf{1.4} [1.4,1.5]
 &\textbf{1.4}[1.4,1.6] & 1.3& 1.0& \textbf{1.4} [1.4,1.4]& \textbf{1.4} [1.3,1.4]\\
\midrule
OU, $\sigma^2_{OU}=64$  & $5.7\times 10^5$& 3.1&  {\textbf{61}} (24,191] & {\textbf{61}} (13,353]& $7.0\times 10^4$& 22& {\textbf{61}} (44,86)& {\textbf{61}} [35,110)\\
 $\rho=.95$  &.64 & .52& {\textbf{.95}}  [.89,.98)& {\textbf{.95}}  [.82,.99)& .71& .57& {\textbf{.95}} (.93,.96]& {\textbf{.95}} (.91,.97)\\ 
 \midrule
OU$+$FBM, $\sigma^2_{1}=2.0$  &\textbf{2.2} & 1.5 & \textbf{1.8} [1.0,3.2] & \textbf{1.8} [.75.4.4]& 1.9 & 2.5 & \textbf{2.0} (1.6,2.4) & \textbf{2.0} (1.4,2.7)\\
     $\alpha=.45$  &1.3 & 1.1 &\textbf{.41} [.31,.55) &.40 (.24,.68) &1.2 &.76 & \textbf{.44} (.41,.48] &\textbf{.44} (.38,.51)\\
 $\sigma^2_{2}=9.0$  &5.6 & 3.2 &\textbf{9.8} [4.0,28] &9.9 (2.1,54) &7.6 &2.1 & {\textbf{9.7}} (7.1,13]  & {\textbf{9.7}}(5.4,17]\\
 $\rho=.85$  &.55 & .43 &{\textbf{.85}} (.74,.93) &{\textbf{.85}} (.61,.96)&.62 &.50 & {\textbf{.85}} (.81,.89) &{\textbf{.85}} (.77,.91]\\

\bottomrule
\end{tabular}
\caption{Parameter estimation for simulation of isotropic processes using different methods. The one closest to the truth is highlighted in bold. Regular videos have 500 $\times$ 500 $\times$ 500, while small videos have size 100 $\times$ 100 $\times$ 100. The brackets give $95\%$ confidence intervals by AIUQ approaches and different types of brackets are due to rounding error. All intervals by AIUQ approaches cover the true parameters given in the first column. 
}
\label{tab:est_simul_theta}
\end{table*}





\section{Simulated studies} 
\subsection{Isotropic processes}
\label{subsec:isotropic_simul_1}

 \begin{figure*}[t]
\centering
\includegraphics[width=\textwidth]{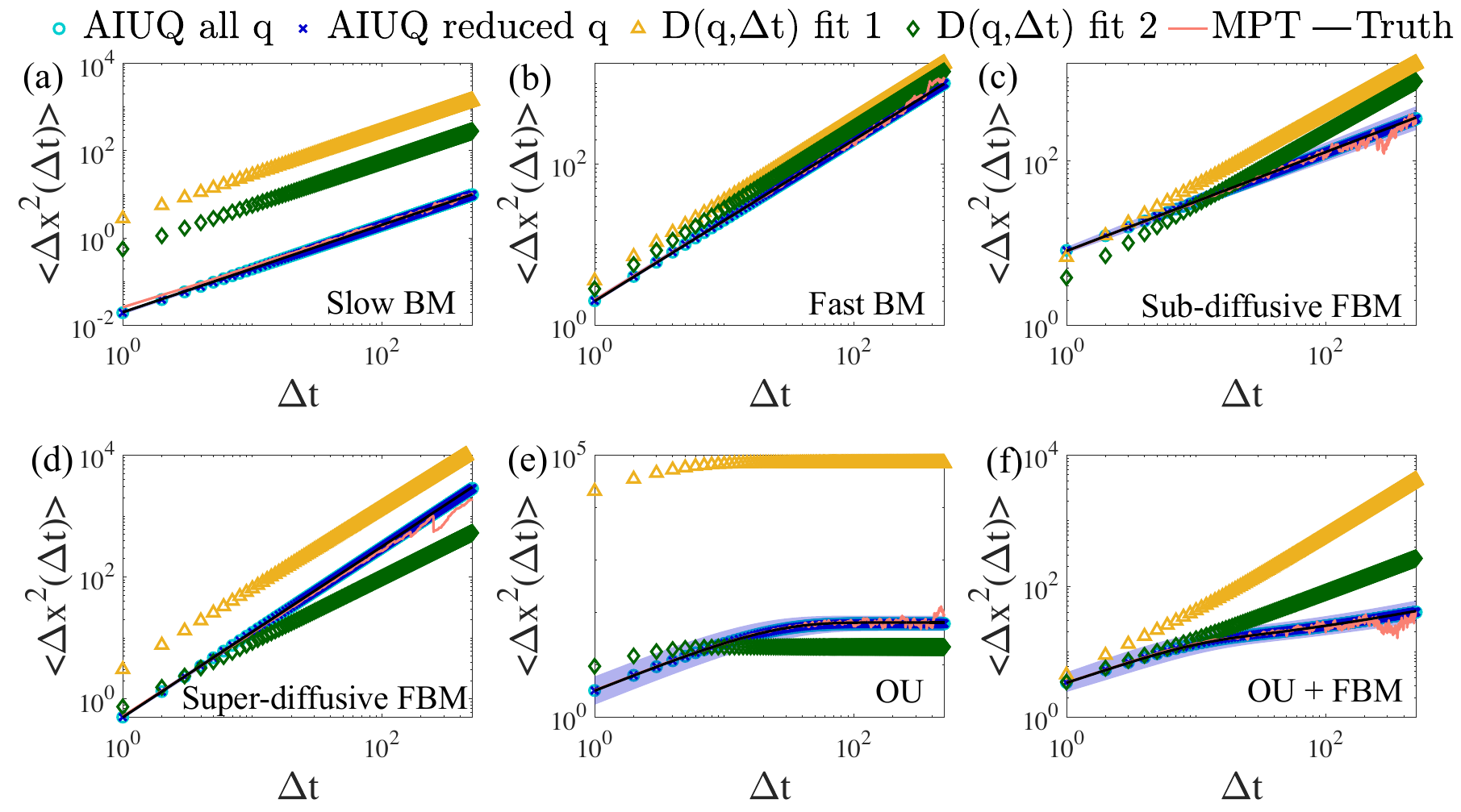}
   \caption{Estimated  mean-squared displacements versus lag time  plotted in log-log space for simulated videos containing the trajectories of $M$ = 50 simulated particles by AIUQ with reduced wave vectors (blue crosses), AIUQ all q (cyan crosses), $D(q,\Delta t)$ fit 1 (purple diamonds) and $D(q,\Delta t)$ fit 2 (yellow triangle). The analysis is for (a) Slow Brownian Motion, (b) Fast Brownian Motion, (c) Fractional Brownian Motion with subdiffusive dynamics, (d) Fractional Brownian Motion with superdiffusive motion, (e) Ornstein–Uhlenbeck process and (f) a mixture of the OU process and FBM.  The solid lines denote the truth. The shaded area denotes the 95\% interval for AIUQ with reduced wave vectors. The truth overlaps with both AIUQ approaches. }
   \label{fig:simulated_studies}
\end{figure*}

We first compare the estimation accuracy of ISF for isotropic processes through simulations. We simulated videos from six different processes: Brownian motion (BM) with two diffusion coefficients, fractional Brownian motion (FBM) with two power parameters corresponding to subdiffusive and superdiffusive processes, respectively,  Ornstein–Uhlenbeck (OU) processes, and a mixture of the  OU process and FBM (OU$+$FBM). 
The ISFs of these processes are provided in Table \ref{tab:ISF_table} in Appendix A. For each process, we test the performance of each approach using both a small-sized video with $100\times 100$ pixels and $100$ time frames, and a regular-sized video with $500\times 500$ pixels and $500$ time frames. The algorithms are applicable to videos with different spatial and temporal lengths. The smaller videos contain fewer frames but retain the  same dynamics.

We include the AIUQ approaches with all wave vectors, reduced wave vectors, two DDM approaches with all wave vectors,  and MPT. 
{The first AIUQ approach uses all wave vectors, which is the default setting in our package.} The second AIUQ approach accelerates the 
 computation  by reducing the first $J_0$ sets (or rings) of wave vectors such that $\sum^{J_0}_{j=1}A_j/\sum^{J}_{j=1}A_j\geq 1-\varepsilon$ and $J_0/J>\beta$ with $\varepsilon=0.001$ and $\beta=0.5$, which leads to at least first $50\%$ of rings of wave vectors being selected. The same choice is applied to all simulations discussed in Sec. \ref{subsec:isotropic_simul_1}-\ref{subsec:anisotropic_simul_3}. 
We found that using a less conservative threshold $\varepsilon=0.01$ and $\beta=0$ leads to around the first $25\%$ of the wave vectors being selected, nonetheless, doing so yields the same estimation for physical parameters in the intermediate scattering function. However, the estimation of the noise parameter may not be as good for small videos. \edit{Here, we reduce the number of wave vectors as an approximation to AIUQ with all wave vectors only to accelerate computation, which is distinct to other approaches that require specifying the wave vector range for estimation.} \newedit{The default setting of our packages is to use all wave vector ranges. 
The two routinely used DDM approaches have been described in Sec. \ref{subsec:comparison_mmle_fitting}}. Finally, in  MPT, we input the known particle radius which enables the algorithm to effectively choose the optimal band-pass filter, whereas other parameters such as search radius and brightness of the center pixel need to be tuned depending on the specific case. 




In Table \ref{tab:est_simul_theta},  the true parameters, the estimated parameters by two DDM approaches, AIUQ with reduced and all wave vectors are provided. The AIUQ approaches yield accurate estimations for regular videos, outperforming the two DDM approaches. AIUQ also yields excellent results even when applied to significantly smaller images. Despite a 125$\times$ reduction in image size, the method yields estimations with an almost indiscernible error. Furthermore, the $95\%$ confidence intervals of each parameter are given in the brackets, which cover the true parameters. 
The significant improvement of the performance by the AIUQ approaches is attributed to its appropriately weighing of information from each wave vector by the marginal likelihood function in Eq. (\ref{equ:prod_density}). Consequently, there is no need to choose a specific range of wave vectors for estimation, other than for the purposes of reducing the computational cost.

\edit{We report estimation of the noise parameter and the truth in Table S3 in the Supplemental Material.
DDM fit 1 produces a reasonably good estimate of the noise parameter, yet the performance in estimating the physical parameters in ISFs is not satisfying. 
\newedit{The estimation of DDM methods can be improved by selecting a wave vector range in a case-by-case manner, and a principled way to select the wave vector range is yet to be found.} We found that even in the diffusive case, identifying an appropriate q-range is non-trivial, as shown in Fig \ref{fig:MSD_pos_record_true_B_40_true_D_0_02} in Sec. \ref{subsec:MMLE} and Fig. \ref{fig:direct_inverstion_dqt} in Appendix B.} 
Avoiding selecting the range of wave vector can automate the scattering analysis of video microscopy, a key feature for such analysis to be 
integrated into any high-throughput experiment. \edit{This has been achieved by the AIUQ approach, where utilizing information in the entire q range leads to good performance.} 

The true MSD, estimated MSD by MPT, two  DDM and AIUQ approaches of six simulated processes with a regular video size are plotted in Fig. \ref{fig:simulated_studies}. Estimation from the two AIUQ approaches and the true MSD curves overlap for all cases, indicating precise estimation of MSD by the AIUQ approaches. Furthermore, the $95\%$ confidence interval by AIUQ (shown as the blue shaded area) is relatively short, in particular for the two BMs, but they cover most of the underlying truth for  MSD.  Lastly, MPT is reasonably accurate for most scenarios with case-specific tuning parameters. However, it has a noticeable discrepancy at small $\Delta t$ for BM with slower dynamics (Fig. \ref{fig:simulated_studies}(a)), and at large $\Delta t$ for almost all processes, whereas AIUQ approaches are robust at small $\Delta t$. Furthermore, we record the root mean squared error (RMSE) between the true MSD and estimated MSD by different approaches in Table S1  
in the Supplemental Material. 
and the AIUQ approaches have the smallest estimation error than other approaches in all scenarios, and in particular, the RMSE by two AIUQ approaches are both  5-10 times smaller than the ones by MPT. This means that if the model is properly selected, the AIUQ approach can be more accurate than MPT in terms of estimating the MSD, and they do not require tuning parameters. \revise{For all methods, estimates for $A(q)$ are presented in Fig. S1}. \edit{Here, we must assume a certain form of the ISF, similar to most studies in DDM. In MPT, it is not necessary to prescribe such a form, though eventually one must connect the estimations derived from MPT to a physical model. Therefore,  extending the AIUQ approach to encompass nonparametric estimation of ISFs  \cite{bayles2017probe,gu2021uncertainty} represents a promising avenue for further exploration. }


The small estimation error by the scattering analysis of microscopy video for the wide range of simulated processes has not been seen before, which is achieved without the need to select wave vectors or tune any parameters. 
The new approach allows for a smaller image sequence with a shorter time interval to be employed, leading to savings in both time and storage, and yields more accurate estimation when videos with a regular size are used.





\subsection{Model selection} 
\label{subsec:selection_simul_2}

 \begin{figure} 
\centering
\includegraphics[width=0.48\textwidth]{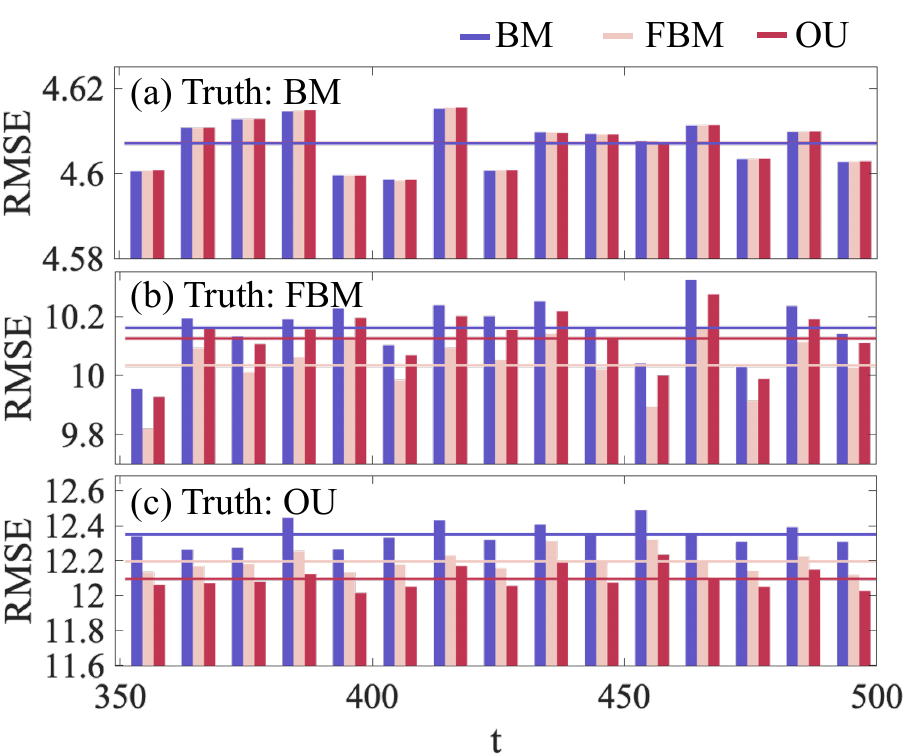}
   \caption{Predictive accuracy as demonstrated by the Root Mean Square Error (RMSE) for (a) Brownian Motion ($\sigma^2_{BM} = 0.02$) (b) Fractional Brownian Motion ($\sigma^2_{FBM} = 8$, $\alpha = 0.6$), and (c) Ornstein–Uhlenbeck process ($\sigma^2_{OU} = 64$, $\rho = 0.5$) trained on the first \newedit{350} lag times. The horizontal lines are the average of the RMSEs of each method. \newedit{The RMSEs are averaged every 10 lag times to improve the readability of the graph. }}
   \label{fig:RMSE_model_selection}
\end{figure}


Conventionally, physical models of intermediate scattering function are selected based on pre-existing knowledge of the dynamical process. Here we introduce two ways for model selection among a few candidate choices for microscopy video from data because of the availability of the probabilistic generative model in Eq. (\ref{equ:SAM}).   

We first perform the model selection by minimizing the predictive error introduced in Sec. \ref{subsec:model_selection}. We simulate microscopy videos of  BM, FBM, and OU processes with the regular size ($500 \times 500 \times 500$). Then we fit three candidate models of ISF for each video using the first  $70\%$ of the time frames, and save the other $30\%$ of the time frames for computing one-step-ahead root means squared errors (RMSEs) for the subsequent time frames, and the average of the RMSEs: $\mbox{AvgRMSE}=\frac{1}{(n-n_0)}\sum^{n}_{t=n_0+1}\left(\frac{1}{N}\sum^{N}_{j=1}( y_j(t)-  y_{pred,j}(t))^2\right)^{\frac{1}{2}} $, where $y_{pred,j}$ is the prediction at the $j$th pixel from Eq. (\ref{equ:y_pred_t}), where $n_0=350$  is the number of time frames used in training the model.

The predictive RMSEs of three models on held-out time frames are plotted as histograms for simulated BM, FBM, and OU processes in  Fig. \ref{fig:RMSE_model_selection} (a)-(c), where each histogram of RMSE is averaged every 10 \newedit{lag times}. The overall AvgRMSEs are shown as the horizontal lines. First, when the underlying process is BM (Fig. \ref{fig:RMSE_model_selection}(a)), the three models have almost the same  RMSE at all different frames. This is because BM is a special case of FBM with $\alpha=1$. The estimated power parameter yields $\alpha_{est}\approx 0.97$, meaning the FBM model approximately reduces to the BM model in this case. 
For the OU process, using the binomial approximation, one has $\sigma^2_{OU}(1-\rho^{\Delta t})\approx \sigma^2_{OU} (1-\rho)\Delta t$, when $1-\rho$ is close to zero. Thus, OU can also approximate  BM relatively well when $\sigma^2_{OU} (1-\rho)\approx \sigma^2_{BM}$. Indeed, we found $\sigma^2_{est,OU} (1-\rho_{est})\approx 0.0196$, a value close to the true sampling model from BM with the parameter $\sigma^2_{BM}=0.020$. When the predictive error of these three models is similar,  the preferred model is BM due to its simplicity, as it contains fewer fitting parameters in its ISF. 
Second, when the true process is FBM (Fig \ref{fig:RMSE_model_selection} (b)), the fitted FBM model consistently yields a smaller predictive RMSE compared to the other two models across most held-out time frames. The AvgRMSE indicated by the pink solid line for FBM remains identifiably lower than the other misspecified models.
 Third, when the true process is OU (Fig \ref{fig:RMSE_model_selection} (c)), the AvgRMSE of fitting an OU, plotted as the red solid line,  is the smallest among the three models. Thus, we correctly select the true sampling models for all three cases using the predictive error. 

Performing model selection with one-step-ahead predictive error requires sequentially predicting intensities on $n-n_0$ time frames, which is time-consuming. Instead, we can compute AIC based on the maximum marginal likelihood value. A model with a smaller AIC is preferred as it indicates a smaller prediction error.  In Table \ref{tab:AIC}, we show AIC by different models in each column. We notice that the AIC of the three models is almost identical when the underlying dynamics follow a BM, as FBM and OU approximate BM with the estimated parameter. 
The BM is preferred when AIC is similar, as it has a smaller number of parameters. Similar to the findings in Fig. \ref{fig:RMSE_model_selection}, we found that the correct model has the smallest AIC, corresponding to the best fit, for FBM and OU. 
Hence, all true models are correctly selected by  the model selection criterion. \newedit{Automated model selection is not well-studied in scattering  analysis of microscopy videos. Therefore, improving the robustness of the methods in complex experimental conditions is of great interest in practice.  }


 \begin{table}[t]
\centering
\begin{tabular}{l|c|c|c}
 \toprule
simulation/fitting & BM & FBM & OU \\ 
 \midrule

 BM  &{\bf 2.2320} & {\bf 2.2320}& {\bf 2.2320} \\ 
\midrule

FBM  & 2.6195 &  {\bf 2.6173}& 2.6189  \\ 
\midrule
OU  &2.6464 &2.6455 & {\bf 2.6445}   \\ 

\bottomrule
\end{tabular}
\caption{AIC of the fitted model is shown in each column. Each row gives one simulation.  The smaller the number indicates a better fit and the smallest value in each row is highlighted in bold. The values are multiplied by $10^8$. 
  }
\label{tab:AIC}
\end{table}


\subsection{Anisotropic processes}
\label{subsec:anisotropic_simul_3}

\begin{figure} 
\centering
\includegraphics[width =0.48\textwidth]{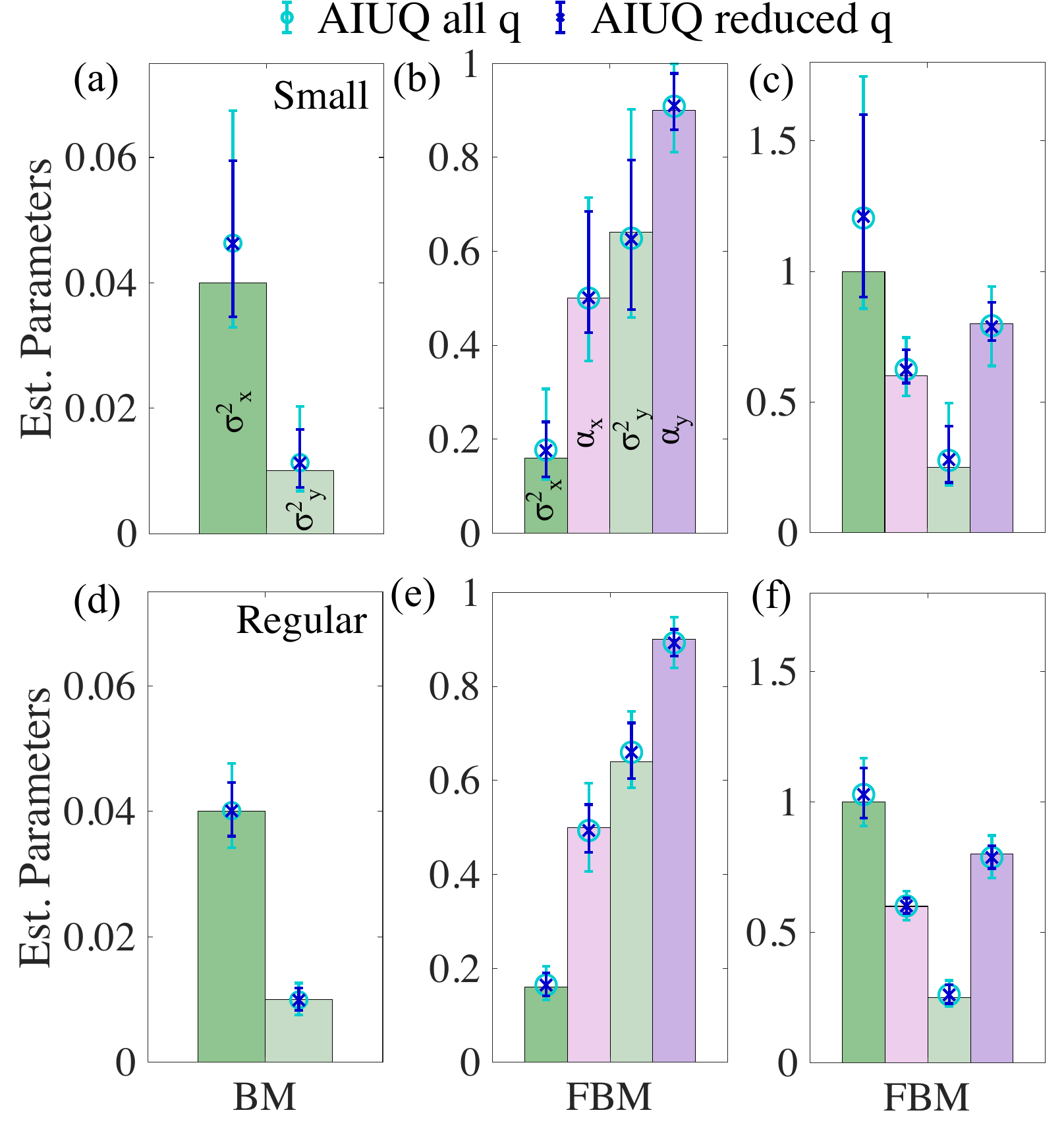}
   \caption{Parameter estimation for anisotropic processes, where the truth is plotted as bars.
   The top row shows small videos (100 $\times$ 100 $\times$ 100), and the bottom row shows regular videos (500 $\times$ 500 $\times$ 500). (a)(d) show the analysis of particles diffusing with two distinct diffusion coefficients along x and y, whereas (b)(c)(e)(f) show fractional Brownian motion with distinct coefficients and exponents.} 
   \label{fig:est_param_bar_graph_anisotropic}
\end{figure}

Here we test the performance of AIUQ approaches for anisotropic processes. We simulate small and regular-sized videos for anisotropic processes from one BM, and two FBMs with different parameter sets for motions along each coordinate. 
The parameters used for simulating these three anisotropic processes are plotted by the color bars in Fig. \ref{fig:est_param_bar_graph_anisotropic}. For BM, the variance of the displacement in the $x$ direction of the process is 4 times as large as the one in the y direction. For the first FBM, the motion in the $x$ direction has a larger variance parameter, but the power parameter is smaller than that in the y direction. For the second FBM, both the variance and power parameters of the motion in the $x$ direction are smaller than the ones in $y$ direction. 

\begin{figure*} 
\centering
\includegraphics[width=\textwidth]{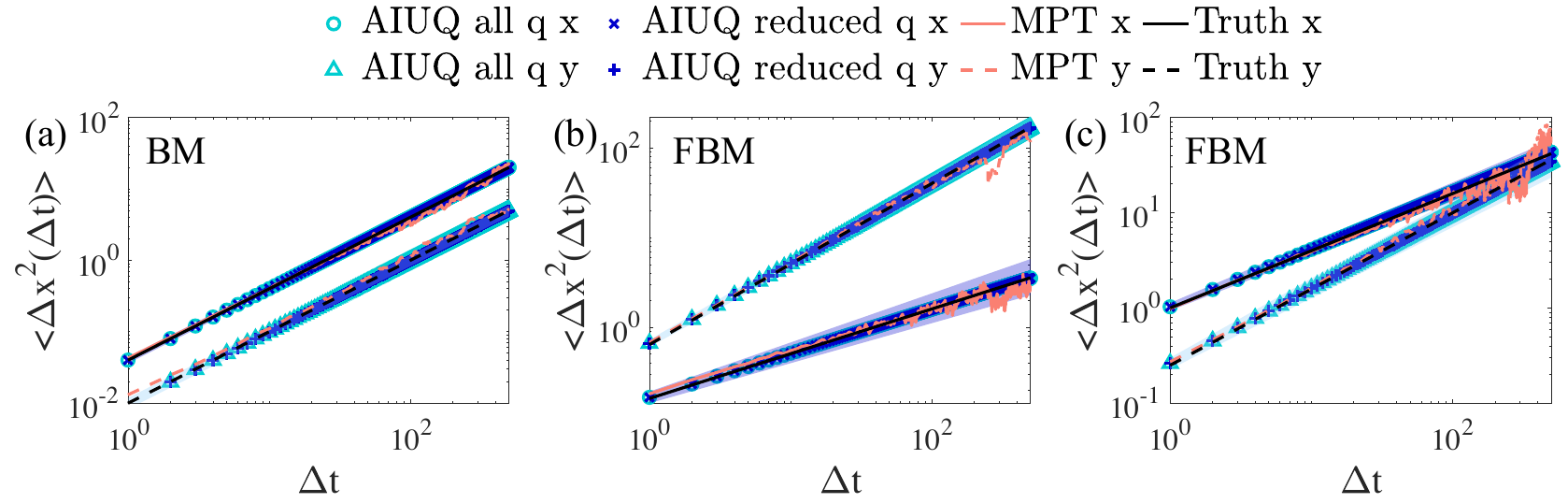}
   \caption{Analyzing simulated studies with anisotropic dynamics. The estimated mean-squared displacements versus lag time are plotted in log-log space based on simulation with $M$ = 50 simulated particles,  by AIUQ (blue crosses), AIUQ all q (cyan circles and triangles), MPT (pink solid lines).  The analysis is for (a) Brownian Motion, (b)(c) Fractional Brownian motion. The shaded area represents the 95\% interval of AIUQ reduced q. The truth overlaps with both AIUQ approaches. }
   \label{fig:simulated_anisotropic}
\end{figure*}

In Fig. \ref{fig:est_param_bar_graph_anisotropic}, we plot the estimated parameters and the $95\%$ confidence interval by two AIUQ approaches for small and regular-sized videos of anisotropic processes. The estimated parameters are reasonably accurate even for most of the small videos, and the estimation error is indiscernible for videos with a regular size. Furthermore, the small $95\%$ confidence interval covers all the true parameters and the interval becomes narrower when the size of the video increases. The lengths of the interval quantify whether the precision of the estimation is satisfied, allowing one to strike a balance between the size of the microscopy video and the precision of the analysis. 

We plot estimated MSD from the AIUQ and MPT in Fig. \ref{fig:simulated_anisotropic}. 
In panel (a), our approach correctly identifies that the diffusion coefficient along x is 4 times as large as that along y.  
Panel (b) shows a subdiffusive case, where the motion along y has both a higher exponent and prefactor, causing the difference of MSD along x and y to grow with lag time. Panel (c) shows another subdiffusive scenario along both axes, where the motion along $y$ has a smaller prefactor yet a larger exponent, causing the MSD along y to grow faster and eventually approach that along x. The estimated MSDs are shown for two AIUQ approaches and MPT. Across all cases, AIUQ accurately replicates MSDs of the simulated dynamics, even when the MSDs in both directions approach one another in the third scenario. 
 Routinely used MPT algorithms \cite{crocker1996methods} typically link particles between two time frames within a pre-specified radius, which may not be optimal for anisotropic processes. Indeed, we found that identifying a suitable set of tuning parameters for MPT is harder for anisotropic processes. 
In comparison, one does not need to tune parameters in AIUQ approaches, and the estimated MSD from AIUQ all overlap with the truth shown in Fig. \ref{fig:simulated_anisotropic}.
 Furthermore, for Fig. \ref{fig:simulated_anisotropic}(a)-(b), MPT has a noticeable discrepancy in estimating the MSD at small $\Delta t$ due to the difficulty in separating small signal from noise, whereas AIUQ approaches are accurate. 
For all processes, the RMSE of estimating MSD by the AIUQ approaches is a few times smaller than the ones by MPT  shown in {Table S2 
in the Supplemental Material.
}.


 DDM approaches were used for analyzing the anisotropic processes \cite{giavazzi2014viscoelasticity,pal2020anisotropic,richards2021particle}, but existing approaches only use a fraction of the transformed data along the x and y coordinates, selected by researchers, and existing DDM packages  do not support estimating anisotropic processes. 
 \edit{Estimation of anisotropic processes has been implemented in our AIUQ packages \cite{AIUQ2024Rpackage,AIUQ2024MATLABpackage}.} 
We utilize information from all wave vectors, yielding better accuracy in estimation and enabling appropriate uncertainty assessment without the need for selecting and additional weighing of the information from different wave vectors. 


We expect that this algorithm will have broad applicability to analyzing a plethora of biological scenarios involving collective cell motion under various conditions such as chemotaxis \cite{theveneau2010collective}, durotaxis \cite{sunyer2016collective}, and haptotaxis \cite{carter1967haptotaxis}, especially for dense settings, typically seen in a confluent cell monolayer. In Sec. \ref{subsec:anisotropic_LC}, for instance, we will introduce the feasibility of analyzing the anisotropic motion for probes embedded in a lyotropic liquid crystal.

\section{Real experimental analysis}
\subsection{Materials and Methods}
Polyvinyl alcohol (PVA, Mw 85,000-124,000, 87-89\% hydrolyzed), dimethylsulfoxide (DMSO), Triton X-100, and disodium cromoglycate (DSCG) salt were purchased from Sigma-Aldrich, and used without further purification. The solutions were weighed and dispersed in ultra-pure water (18.2 M$\Omega$-cm). 4-arm polyethylene glycol (PEG) polymers, terminated with primary amide (NH2) and succinimidyl glutarate (SG) groups were purchased from JenKem Technology USA. 

Probes (FluoSpheres, yellow-green, carboxylate-modified microspheres) of different sizes ($2r_p$ = 100 nm, 200 nm, and 1 $\mu$m) were purchased from Thermo Fisher. All samples were prepared by filling a square capillary (0.10 mm x 1.0 mm x 0.09 mm, Friedrich \& Dimmock Inc.), sealing on two ends, and securing onto a glass slide with UV curable glue (Norland Optical Adhesive), to minimize convection due to leaking and evaporation.

The samples \newedit{are} imaged using a Zeiss Axio Observer 7 microscope in fluorescence mode using a Colibri 7 light source,  and standard GFP filter sets. The images are captured with a 20x objective, with a numerical aperture of 0.8.  A typical image size is 512 $\times$ 512 pixels with a pixel size of 0.29 $\mu$m/pixel and n = 500 time steps with a step size of 0.0309 seconds.

\newedit{We have simulated 2D motion for simplicity, but the probe movements are intrinsically 3D in real experiments. For these analyses, the motion of the particle is projected onto the plane of the focus and only 2D trajectories are obtained and analyzed \cite{gardel2005microrheology}.  There
are cases for which this simplification does not work and the observed dynamics are affected by axial motion, such as shown in  \cite{lu2012characterizing}. Nonetheless, in all the cases shown here, the axial
displacement is either negligible, or the motion is isotropic in all three dimensions so a 2D treatment of the 3D scenario is warranted.}

\subsection{Diffusive motion with different particle size and number density}
\label{subsec:diffusive_real}
 \begin{figure*} 
\centering
\includegraphics[width=\textwidth]{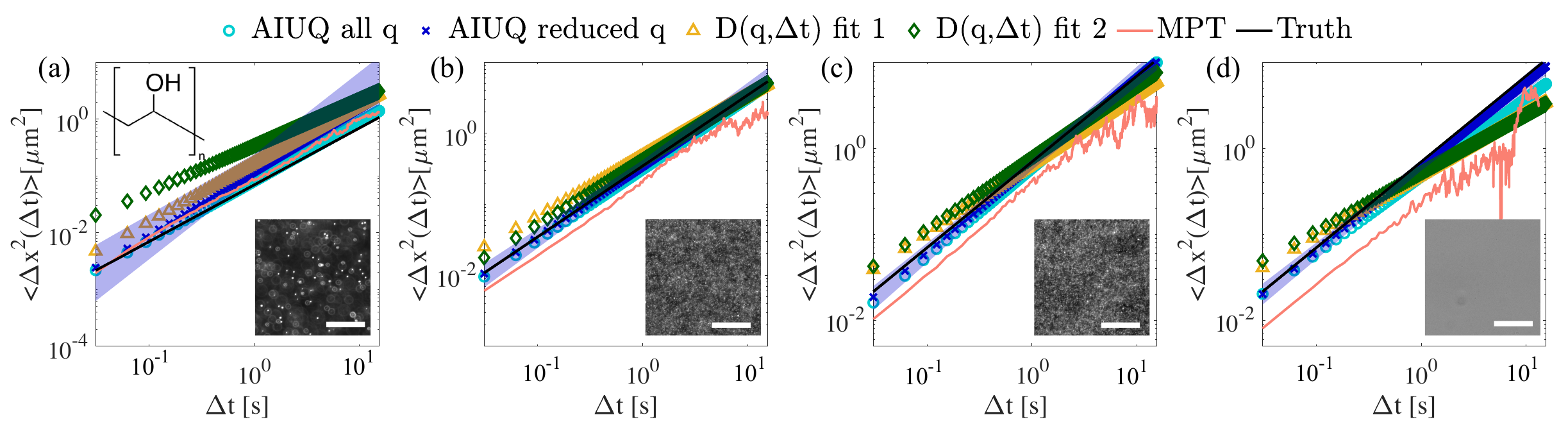}
   \caption{The diagram shows the mean-squared displacement of probes in a 4 wt\% PVA solution, and the molecular structure is also shown in the inset in (a). \newedit{(a-c) consist of fluorescent images whereas (d) consists of bright field images. }(a-d) The mean squared displacement against the lag time is plotted in log-log space for different embedded particle sizes: (a) $2r_p = $ 1 $\mu$m, (b) $2r_p = $ 200 nm, (c) and \newedit{(d)} $2r_p =$ 100 nm\newedit{, where the image sequence was captured in bright field mode in (d).} The inserts contain snapshots illustrating the typical probe size and concentration.  Each image is 150 $\mu$m $\times$ 150 $\mu$m. The scale bars are 50 $\mu$m. AIUQ with all q is shown in cyan circles, AIUQ is shown in blue crosses, \edit{$D(q,\Delta t)$ fit 1 and 2 are shown in yellow triangles and green diamonds, respectively}. The solid pink line denotes tracking by MPT. The solid black line denotes literature values, where viscosity $\eta \approx$ 0.025 Pa$\cdot$s. The blue, shaded region denotes the confidence interval using AIUQ reduced q. }
   \label{fig:diffusive_2sizes_exp}
\end{figure*}

Polyvinyl alcohol (PVA) finds extensive use in spinning applications, making its rheological properties pivotal for efficient processing. In this context, we measure the viscosity of a 4(w/v)(\%) of PVA in water, which is in the dilute regime. The solution behaves as a Newtonian fluid with a viscosity of $\eta \approx$ 25 mPa$\cdot$s per manufacturer information. Here, we aim to demonstrate that the scattering analysis of microscopy by the AIUQ approach can overcome potentially challenging scenarios for MPT, such as optically dense samples due to large particle numbers or fast dynamics leading to unidentifiable switching of particle positions. \edit{The standard DDM was shown to outperform MPT in an optically dense system \cite{bayles2017probe}. However, this was only possible when a specific wave vector range was selected. A method using all wave vectors (i.e. not selecting q) has not been demonstrated before. }

The MSD of a diffusive process can be related to viscosity $\eta$ by the Stokes-Einstein equation \cite{einstein1905molekularkinetischen}:
\begin{equation}
    \langle \Delta x^2(\Delta t)\rangle = \frac{2k_B T_a}{3\pi \eta r_p}\Delta t,
\end{equation}
where $k_B= 1.38 \times 10^{-23} J/K$ is the Boltzmann constant, $T_a$ is the absolute temperature, $r_p$ is the radius of the particle. 
Keeping $T_a$, $\eta$ constant, the slope of $\langle \Delta x^2(\Delta t)\rangle$ versus $\Delta t$  decreases proportionally to increasing particle radius $r_p$. 
We prepare several \newedit{samples} of the same composition but adding particles of different sizes $2r_p = $ 1 $\mu$m, 200 nm, and 100 nm in each one. All \newedit{fluorescent image sequences were captured at} particle volume fraction $\phi$ = 1$
\times 10^{-4}$, but the number densities of the 200 nm and 100 nm particles are much higher, as shown in the insets in Fig. 
\ref{fig:diffusive_2sizes_exp}. \newedit{The bright field image sequence was captured at particle volume fraction $\phi$ = 2$
\times 10^{-3}$.}

We compare MPT,  AIUQ approaches and two fitting approaches based on image structure functions. \edit{For AIUQ approaches, we show the performance based on all q or reduced q to accelerate the computation.} For all real experiments in Sec.~\ref{subsec:diffusive_real}-Sec.~\ref{subsec:anisotropic_LC}, the smallest $J_0$ such that   $\sum^{J_0}_{j=1}A_j/\sum^{J}_{j=1}A_j\geq 1-\varepsilon$ with $\varepsilon=0.99$ is used as a default choice of the wave vectors for reducing the computational cost in the AIUQ with reduced q approach. Such choice leads to $J_0\approx 0.25 J$, which reduces the computational cost by more than 10 times compared to AIUQ on all $q$, suitable for scalably computing a large number of experimental data, especially for those discussed in Sec. \ref{subsec:gelling}. For the three sets of experiments discussed here, we assume the FBM model to parameterize ISF in AIUQ approaches and fitting image structure function approaches instead of BM to test whether the model can identify the Newtonian behavior of the fluids,  
as well as validate that
other potential factors, such as drifts, do not have a large impact on the results.   

\edit{As shown in Fig. \ref{fig:diffusive_2sizes_exp}, parameter estimation by fitting $D(q,\Delta t)$ is less accurate than the by AIUQ approaches, based on either all $q$ or reduced $q$, compared to the reference value (black curves). This result highlights that the wave vector range needs to be carefully selected when fitting image structure function with conventional DDM, but this problem is resolved using AIUQ, which uses the maximum marginal likelihood estimator.} 
For nanoparticles with $2r_p = $ 100 and 200 nm in panels (b)-\newedit{(d)}, 
which are all below the diffraction limit, 
 MPT systematically underestimates the particle movements compared to the truth, due to higher concentrations and faster dynamics. MPT computes the ensemble MSD by linking the trajectories and averaging the displacements from all particles. When dealing with an optically dense sample, the probability of erroneously linking two nearby particles increases.  
 Furthermore, the progressively worsening performance of MPT is attributed to averaging fewer particle steps towards the larger $\Delta t$'s. We note that quite often, MPT only utilizes the first 10-20\% of the $\Delta t$'s, requiring  substantially larger time range to be measured. \newedit{Furthermore, Fourier-based algorithms are better for MPT at analyzing data sets with a low signal-to-noise ratio, such as the brightfield image shown in panel (d), where particles are unresolvable by the eye. }
\revise{As further verifications, estimates for $A(q)$ are presented in Fig. S2, and observed image structure function $D(q,\Delta t)$ and the estimated $D(q,\Delta t)$ using either AIUQ or DDM are shown in Figs. S3 and S4}.
 
The estimated MSDs by  AIUQ approaches are closer to the reference values than MPT for scenarios in  Fig. \ref{fig:diffusive_2sizes_exp}(b)-\newedit{(d)}, 
as AIUQ overcomes the difficulty of separating and linking a large number of particles. 
Optically dense systems are not uncommon in experiments. 
In some experiments, for instance, larger particles are incompatible with the system,  when the density and viscosity of the continuous phase are both low - probe particles tend to sediment unless they are small enough to be dispersed by Brownian forces \cite{gu2021uncertainty}. The second scenario is when higher moduli are expected and thus using small probe particles is necessary to produce detectable displacements. A third scenario is when the system already contains probes, such as tracking phase-separated regions inside organelles \cite{brangwynne2009germline}, where the size  of the particles cannot be adjusted. AIUQ approaches are well-suited for analyzing optically dense samples without the need for \newedit{introducing fluorescent particles}, specifying wave-vector ranges, or tuning parameters. 

Finally, all approaches seem to slightly overestimate MSD at a large $\Delta t$ for the experiment with $2r_p$ = 1 $\mu m$ particles shown in Fig. \ref{fig:diffusive_2sizes_exp}(a). This is because small drift disproportionally impacts the larger probes, which move more slowly under the thermal fluctuation. Using smaller particles such as those in Fig. \ref{fig:diffusive_2sizes_exp}(b)-(c) can mitigate these impacts. Another way is to  model the drift from ISF  and integrate it into AIUQ estimation.  


\subsection{Automated estimation of gelling point of a perfect network}
\label{subsec:gelling}

\begin{figure*} 
\centering
\includegraphics[width=0.95\textwidth]{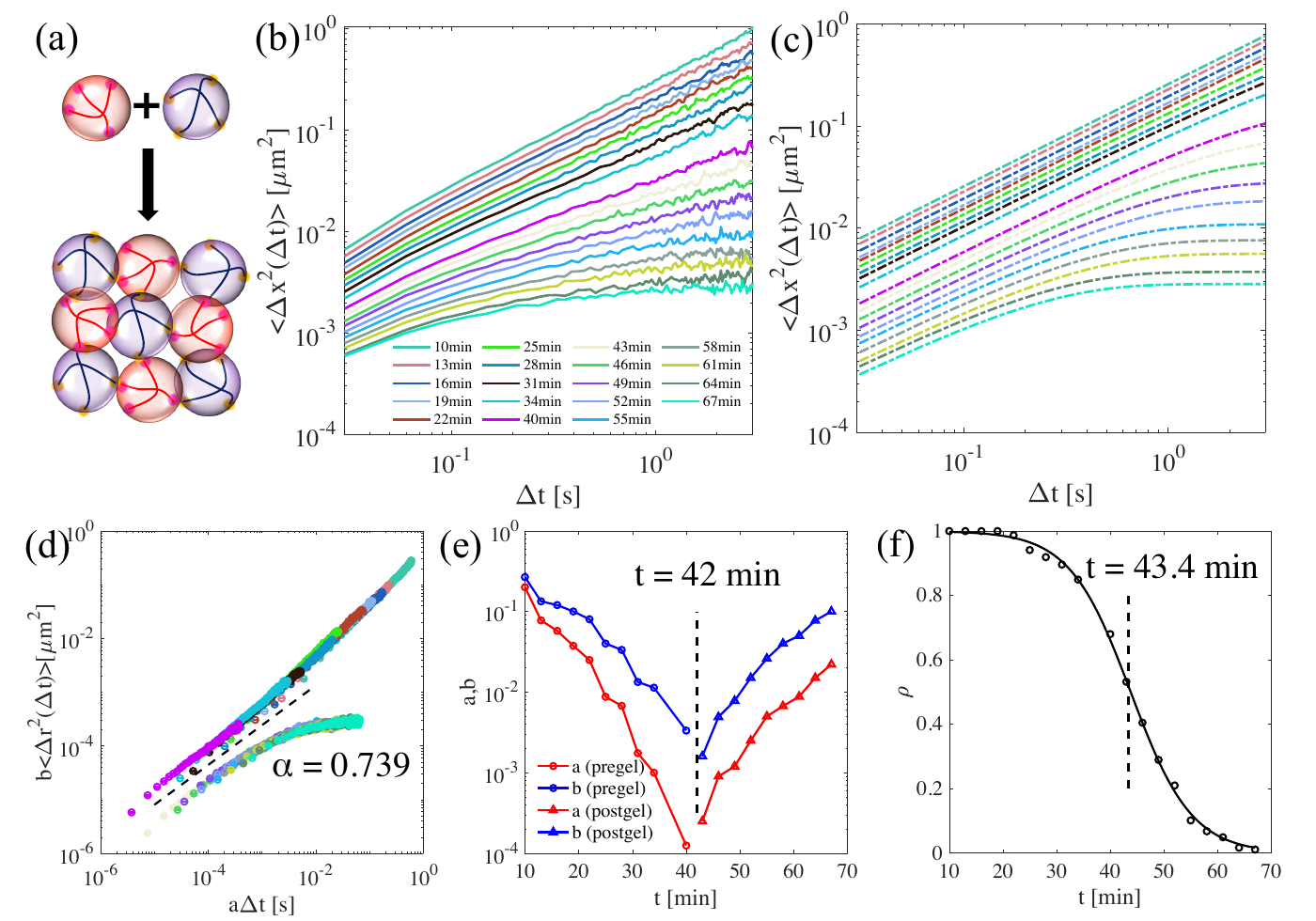}
   \caption{(a) Schematics showing the reaction between tetraPEG-SG and tetraPEG-NH2. (b) Multiple particle tracking of the time series ensemble mean-squared displacement of the probes embedded in the mixed tetra-functional PEG plotted against lag time $\Delta t$ in log-log space. (c) AIUQ analysis of the same data as (b), denoted by the same color but with dotted lines. (d) Superposition of the data in (b) using MPT data. (e) Shift factors $a$ (red) and $b$ (blue) for pregel (circles), and postgel (triangles). (f) Estimated $\rho$ parameters in the OU process from a sequence of videos at different times denoted by black circles. The black solid line denotes the fit from a generalized logistic curve where $R^2_{adjusted} > 0.99$.
   } 
   \label{fig:solgel_exp}
\end{figure*}

A sol-gel transition is one where a solution is gradually transformed into a viscoelastic network, with progressively more solid-like characteristics. This phenomenon is observed in many naturally occurring polymers, such as collagen \cite{shayegan2013microrheological}, protein solution \cite{meleties2022high}, peptides \cite{larsen2008microrheology} and silk fibrin \cite{martineau2022engineering}. Probing the timescale of the  viscoelastic 
properties of the biopolymers can facilitate their use in reconstituted bioscaffolds.

Similarly, synthetic hydrogel materials with customized architecture can assemble into networks mimicking the mechanical properties and structure of their biological counterparts. Because of polyethylene glycol (PEG)'s inert properties, they are widely used as cell substrates \cite{schultz2015measuring}, tissue scaffolds \cite{zhu2010bioactive}, and for cell encapsulation applications \cite{mcglynn2021human}. 
Time-cure or time-concentration superposition analysis of microrheology experiments was introduced by Larsen and Furst \cite{larsen2008microrheology}. 
To date, the tried-and-true method to determine gelling time from microrheology data is mostly done by manually
shifting the MSD curves \cite{martineau2022engineering,schultz2013monitoring,mcglynn2020multiple}. 
 


 
 

The shifting is carried out by multiplying the MSD curves and the $\Delta t$ by coefficients $a$ and $b$, determined by visual inspection. 
Once done, the MSD curves are collapsed into two distinct branches, one pre-gel and one post-gel. 
Each dataset (for a particular gelling time) should exhibit some  overlap with the subsequent one. The MSD at the gelling point is a critical curve between these two branches, which exhibits power-law-like behavior.
The critical gelling exponent is defined as $\alpha = \frac{d\log\langle\Delta x^2(\Delta t)\rangle}{d\log\Delta t}$ \cite{furst2017microrheology,larsen2008microrheology}, where $\alpha$ is constant over all $\Delta t$, identical to the FBM model. 
Different systems experience gelation with distinct critical power-law exponents, as noted in \cite{suman2020universality}. These exponents can span from as high as $\alpha = 0.88$ \cite{lue2008investigation} to as low as $\alpha = 0.16$ \cite{schultz2013monitoring}.

 A recent approach \cite{lennon2023data} aimed to automate the process, using MSD versus $\Delta t$ curves estimated by MPT. However, these MSD curves are non-smooth and therefore  extra efforts are required to smooth the MSD curves before implementing the superposition.  DDM has also been utilized to explore gelation behaviors, \edit{though manually superposing the MSD curves is still needed.  }
 In \cite{martineau2022engineering},  DDM is used as an initial screening step and the authors identify gelation when displacement falls below a threshold, at which point it is challenging to analyze subtle displacements using DDM, \edit{whereas  the pre-specified wave-vector range can affect the estimation.}   
In \cite{meleties2022high}, the authors conducted MSD fitting using both MPT and DDM. Then they extracted the log-slope of MSD curves from MPT and subsequently fit a logit function to these slopes to identify gelation. 

%

 Here, we study the gelation of SG:NH2 = 1:1 tetraPEG mixtures. These functional groups react stoichiometrically to form highly regular networks (Fig. \ref{fig:solgel_exp}(a)).  Stock solutions for both 4-arm PEG polymers (100 mg/mL) were prepared to ensure that the sample was well-mixed. Due to the spontaneous nature of the SG-NH2 reaction, 
 the reaction rate is entirely controlled by solution concentration, hence, we choose this system to benchmark the gelation time, which can be compared to previous study \cite{parrish2020nanoparticle} with a reported gelation time of $\sim$ 44 min for a concentration of 20 mg/mL polymer. To counter hydrolysis, which is known to occur for SG groups dispersed in water, the 4-arm PEG-SG stock solution was prepared in DMSO. 
We first plot the estimated MSD by MPT in Fig. \ref{fig:solgel_exp}(b), which shows that the probe movements are initially diffusive, but at longer $T$, the probe motion becomes subdiffusive and the onset of a solid plateau begins to appear at longer $\Delta t$'s, meaning the particle becomes caged within the developing network. 

We superpose the MPT results by multiplying $\Delta t$ by a set of time shift factors $a$, and MSDs by shift factor $b$ to construct the master pre- and post-gel curves in Fig. \ref{fig:solgel_exp}(d).  By plotting $a$ and $b$ against time in Fig. \ref{fig:solgel_exp}(e), we approximate the gelation time to be $\approx$ 42 minutes, in good agreement with the reported gelling time $T_{gel}=44$ min \cite{parrish2020nanoparticle}. 
However, it becomes difficult to classify which branch the shifted MSD curve should fall close to the critical gelling curves, so the curves corresponding to the smallest shift factors are typically chosen, as shown in Fig. \ref{fig:solgel_exp}(e). Then the gelling point is defined as the time point when shift factors of the pre-gel and post-gel classes diverge, leading to the largest shift or changes between the curves.
 
Next, we demonstrate using AIUQ to automate the gelling point determination.  
The viscoelastic solid can be modeled by an OU process,  with $\mbox{MSD}(\Delta t)=\sigma^{2}_{OU}(1-\rho^{\Delta t})$, which can capture the plateau and the reducing gradient of the MSD curve at large $\Delta t$'s. 
As the number of experiments is large, we use AIUQ with reduced $q$ for estimating the ISF. The estimated MSDs from the AIUQ approach with an OU model are shown in Fig. \ref{fig:solgel_exp}(c), which resemble curves from MPT curves from Fig. \ref{fig:solgel_exp}(b). 
The two parameters from the OU models, $\rho$ and $\sigma^2_{OU}$, determine the shape of the MSDs. These estimated parameters can be further processed to deduce the gel point of the material. 
 In Fig. \ref{fig:solgel_exp}(e), we plot the estimated parameter $\rho$ from the OU model for each experiment. The estimated $\rho$ gradually decreases, due to the caging effect from the network that traps particles at longer $\Delta t$'s. Before gelling, the absolute change of estimated $\rho$ increases, and the absolute change decreases after gelling. The rapid drop of $\rho$ from 1 to 0 indicates the sol-to-gel transition. To find the gelling point, we fit a generalized logistic function, $\rho(t)=\exp(-c_1(t-c_2))/(1+\exp(-c_1(t-c_2)))$, where $t$ is the reaction time,  $c_1$ and $c_2$ are determined by minimizing a $L_1$ loss with respect to $\rho$. The $L_1$ loss is more robust than the $L_2$ loss, and here both give similar estimations. The fit is shown by the 
 black solid line, which characterizes the change of $\rho$, and defines the gelling time to be the time point with the largest intermediate change in $\rho$. The inflection point is found to be $43.4$ minutes using this fit, which is similar to the estimate $44$ minutes in the previous study \cite{parrish2020nanoparticle}. 

To extract the critical gelling exponent, we again compare two approaches. For the MPT approach, we plot the log shift factors $\log a$, $\log b$ against $\log$ of the extent of reaction $c$, defined as $c = \frac{\vert{t-t_{gel}}\vert}{t_{gel}}$, where $t_{gel}$ is the estimated gelling time, hence $c$ determines distance to the gelling point \cite{larsen2008hydrogel}. Given $ a \propto c^{-\alpha_a}$, and $ b \propto c^{-\alpha_b}$, the scaling exponent is the ratio of these two exponents $\alpha = \alpha_a/\alpha_b$. This way, we obtain the scaling exponent for the 4-arm PEG-NH2 and 4-arm PEG-SG system to be $\alpha$ = 0.739. The slope is plotted in Fig. \ref{fig:solgel_exp}(d) for reference. A second way is similar to what has been described in \cite{meleties2022high}, the authors fit a logistic function to the relaxation exponent of the MSD. The same information can be obtained in AIUQ by fitting the fractional Brownian motion to  the MSD curves closest to the gelling point, and we found that the critical exponent of $\alpha$ = 0.74 
for the estimated gelling  point t = 43.4 min. Thus, the scattering analysis of microscopy by the AIUQ approach can be used to automatically extract the critical quantities of gelling systems. 


The AIUQ approach extends the boundary of previous techniques in the post-gel branch of the MSD for systems undergoing gelation.
Compared to MPT and conventional curve shifting for superposition by hand,  AIUQ lifts the hurdles of strenuous analysis for a large number of videos by providing an automated estimation of gelling time,   \edit{without the need of specifying wave-vector range or superposing the curves.} 
Thus the new tool can be deployed for \edit{automatically estimating the gelation time} by a sequences of microscopy videos, at a given formulation and experiment condition (\textit{e.g.} temperature, pH, etc).
Along with high-throughput experiments, this data-processing technique can be integrated with Bayesian optimization and active learning approaches \cite{shahriari2015taking,fang2022reliable} to optimize the compositions or material designs to achieve ideal gelling time and properties.  


\subsection{Anisotropic diffusion in lyotropic LC}
\label{subsec:anisotropic_LC}
 \begin{figure*}[t] 
\centering
\includegraphics[width=\textwidth]{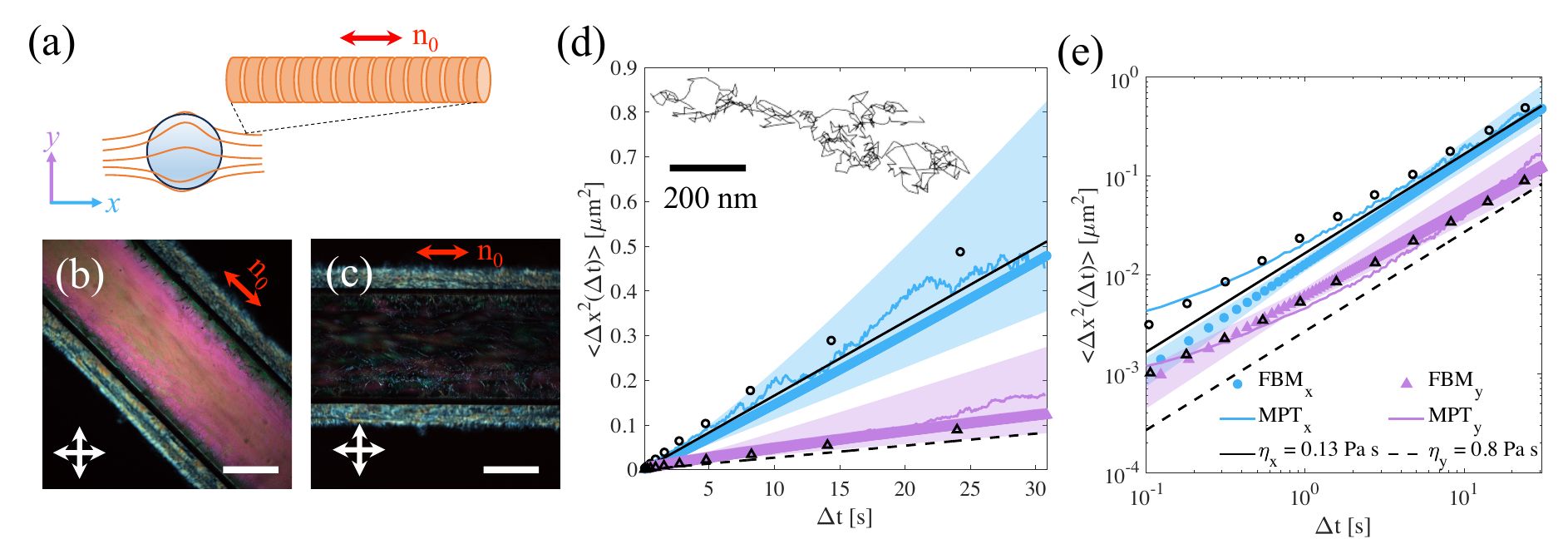}
   \caption{Anisotropic diffusion in DSCG. (a) Schematics of a probe particle moving amongst assembled stacks of DSCG, not drawn to scale. (b-c) Crossed-polarizer images when the channel is aligned either 45$^o$ in (b) or parallel to the alignment direction of the channel in (c). The double-sided arrows denote the direction of the polarizer and analyzer. (d-e) Method comparisons between AIUQ fitting (filled symbols) and MPT tracking (solid lines) of the image stack are presented in (d) lin-lin plot and (e) log-log plot. \edit{Open circles and triangles denote data reported in \cite{gomez2016two}. Black solid and dashed lines denote diffusive trends plotted from viscosity values reported from the same paper.} The inset in (d) shows an example particle trajectory. The shaded region denotes 95\% confidence interval estimated using AIUQ reduced q.}
   \label{fig:anistropic}
\end{figure*}

Anisotropic motion is common in soft and biological systems containing passive particles or active agents, often due to the presence of anisotropic microstructures. An example can be seen in particles within an LC continuous phase. This leads to anisotropic particle motion, aligned either parallel or perpendicular to the orientation of the liquid crystal molecules, which is also called the ``director'' field. 
Here, we study particles dispersed in lyotropic LC disodium cromoglycate (DSCG), one of the most common lyotropic chromonic liquid crystals, where the molecules self-assemble into rod-like structures (Fig. \ref{fig:anistropic}(a)). \edit{While previous work extracted anisotropic MSD by particle tracking \cite{turiv2013effect,martinez2018brownian} and two-point microrheology \cite{gomez2016two}, no method has yet demonstrated successful extraction of these quantities using Fourier-based approaches.}

We disperse probes $2r_p = $ 200 nm  at volume density $\phi$ = 8$\times 10^{-5}$ in 16 wt\% DSCG solution. This concentration is chosen as its iso-nematic transition temperature is above room temperature so no isotropic to nematic transition occurred, as the iso-nematic fronts have been found to trap particles and cause aggregation. Given a density of 1.55 g/cm$^3$ for pure DSCG \cite{koizumi2019effect}, the solution density is 1.088 g/cm$^3$ which is similar to that of polystyrene particles at 1.055 g/cm$^3$ according to the manufacturer, hence these particles are neutrally buoyant. A small amount of surfactant (Triton X-100, 0.015 wt\%) is added to prevent aggregation of the probes \cite{martinez2018brownian}. The capillary force upon filling is sufficient to ensure good alignment of the DSCG solution \cite{gomez2016two}. 
   Uniform nematic alignment was obtained this way, as shown by the polarized optical microscopy images in Fig. \ref{fig:anistropic}(b)-(c).  In these images, regions aligned parallel or perpendicular to the polarizer or analyzer direction (denoted by double-sided arrows) appear dark, while other orientations appear bright.

An example of the anisotropic particle trajectory is shown in Fig. \ref{fig:anistropic}(d) inset. A probe particle moves predominantly along the x direction (sky blue) compared to y (purple). 
The previous MSD by particle tracking results from previous work \cite{gomez2016two} is plotted by \newedit{open circle and }triangle symbols in Fig. \ref{fig:anistropic}. The authors also determined the viscosities parallel and perpendicular to the alignment direction, which are plotted as solid and dashed lines in Fig. \ref{fig:anistropic}.
Since the log-slope $\alpha = \frac{d\log\langle\Delta x^2(\Delta t)\rangle}{d\log\Delta t}<1$, we use the FBM model in the AIUQ approach. 

MSDs by MPT and AIUQ are shown in the lin-lin plot in Fig. \ref{fig:anistropic}(d) and then again in log-log in Fig. \ref{fig:anistropic}(e). 
\edit{As the computation cost for anisotropic processes is higher than isotropic processes, due to the larger number of parameters, and distinct ISFs at each wave vector,} we explore this kind of behavior with a reduced number of $q$'s, but the estimation from AIUQ with all $q$ is similar. 
Both MPT and AIUQ effectively capture large distinctions in diffusion along with direction, and the results are similar to the values reported in \cite{gomez2016two}, as shown in  Fig. \ref{fig:anistropic}(d). 


When examining the log-log plot of MSD, the MSD by MPT  suggests a subdiffusive region at a small $\Delta t$. 
\edit{In a similar system, a subdiffusive-to-diffusive transition was observed at the time scale of twist relaxation, which is on the order of $\Delta t \approx$ 100 s \cite{martinez2018brownian}, but with much larger particles ($2a\approx$ 7 $\mu m$).} 
The authors postulate that subdiffusive behavior can be attributed to the restoring forces on the particle from incurred elastic free energy costs, partially offsetting the displacement from thermally driven fluctuations. \edit{Given that the model used in AIUQ for our case is FBM, we constrain the log-log plot of the MSD to be linear. As noted in \cite{gomez2016two}, physical models that can explain anistropic subdiffusive behaviors in nematic liquid crystal has yet to be constructed. This indicates that a model-free or nonparameteric of ISF is appealing when the physical model of ISF is unavailable, whereas these approaches  typically require a specified wave-vector range for inverting the image structure function in DDM  \cite{bayles2017probe,edera2017differential,gu2021uncertainty}. Whereas utilizing the plateau or variability of the image structure curves can be helpful for identifying the wave-vector range, it is of interest to integrate the model-free ISF in AIUQ approaches for inverse estimation.     }

\section{Discussion}
Minimizing a loss function is often required for estimating parameters in physical and machine-learning approaches. Selecting the loss function and data regime or transformation of parameters implicitly reflects one's  assumptions of data. \edit{To mitigate bias}, we introduce a principled way to find a generative model for approaches that minimize a loss function in two steps. First, we construct a probabilistic model of the untransformed data from the beginning and show that a loss-minimization approach is equivalent to a common statistical estimator of the generative model.  Second, we  integrate out the random component of the model to derive a more efficient estimator, such as the maximum marginal likelihood estimator herein, which naturally \edit{aggregates} the information from different regimes of the transformed data.  
A generative model offers a probabilistic mechanism that allows for the derivation of the asymptotically optimal estimator, and the propagation of uncertainty from the initial stages of data analysis, which we term \textit{ab initio} uncertainty quantification (AIUQ). 


Here, \edit{we use} DDM \cite{cerbino2008differential,giavazzi2009scattering} \edit{as an example to illustrate how building a probabilistic  model can connect the original estimator from standard DDM and to improve conventional loss-minimization methods through the maximum marginal likelihood estimator.} We show that the estimation in DDM analysis is equivalent to minimizing the temporal variogram of the projected intensity in the reciprocal space based on our probabilistic model.
Compared to tracking-based algorithms, DDM eliminates the need to track individual features, but selecting a range of wave vectors is still typically required to minimize the loss function, which can differ in a case-by-case manner.  
With the probabilistic model of data, we derived the maximum marginal likelihood estimator of the parameters, 
which optimally weighs information at each wave vector, removing the need for selecting a wave-vector range to analyze. 
By evoking the generalized Schur algorithm for Toeplitz covariance and further reducing data by truncating the high-frequency wave vectors, we can accelerate the computation by around $10^5$ times for a microscopy video of regular size, compared to directly computing the likelihood function, allowing almost real-time analysis. \edit{We have implemented the AIUQ approaches, and shared them as publicly available software packages available in R and MATLAB \cite{AIUQ2024Rpackage,AIUQ2024MATLABpackage}.}

Through a variety of simulated and real experiments of both isotropic and anisotropic processes, \newedit{imaged in both fluorescent and bright field modes, }
we found that the tuning-free AIUQ approach achieves a high estimation accuracy of model parameters and MSDs which were not seen before, justifying the efficiency in integrating the information at different wave vectors by the likelihood function. The $95\%$ confidence intervals of  MSDs from the AIUQ estimation are typically narrow yet they cover the true MSDs for most $\Delta t$'s, indicating precise uncertainty quantification.  \newedit{Similar to DDM, our algorithm is robust even in scenarios with small signals on a bright background.} Furthermore, using either the maximum likelihood value or predictive error, our method is able to correctly identify the true model amongst a few possible candidates using imaging data. This aspect had not been previously explored within this context.  
The key aspect of the AIUQ approach is that it removes the need for selecting wave-vector range and weighing information on each wave vector in a case by case manner. This feature is particularly useful to be integrated into high-throughput experiments for automatically determining the gelation time at various experimental conditions.

\edit{We outline a few future direction that will overcome the limitation of the AIUQ approaches in the implemented packages. First,  we assume that a physical model of the process and equivalently the parametric form of the ISF is known.  
However, this can be restrictive for scenarios where the underlying model governing the dynamic process is unknown. \newedit{DDM is also used without fitting \cite{bayles2017probe,edera2017differential} by  inverting the image structure function separately to estimate the mean square displacement at each $q$, but it still requires the selection of a range of admissible $q$'s.}  
 In particular, model-free approaches in DDM  proceed by directly inverting the ISF at a selected wave-vector range separately at each lag time \cite{gu2021uncertainty}, inevitably truncating the lag time range that cannot be reliably analyzed. 
\newedit{A robust way for estimating the model-free ISF is of great interest}. Rather, \newedit{by adopting the AIUQ framework,} the ISF approximated by MSD can be optimized using the likelihood function in Eq. (\ref{equ:prod_density}) to efficiently weigh all information at different wave vectors and lag time.  Using the likelihood function in Eq. (\ref{equ:prod_density}) can substantially improve the efficiency of a similar approach in \cite{gu2021uncertainty} by minimizing the loss based on the image structure function,  
which will facilitate the connection of MSD data to frequency-dependent viscoelastic moduli using the GSER \cite{mason2000estimating}.
Second, We find that a small drift is \newedit{crucial to model} for slow dynamics, such as the motion close to the gelation point. \newedit{The AIUQ framework allows one to include the effect of drifts in modeling ISF, and jointly estimate the intrinsic thermal fluctuation  and drifts due to external conditions}. 
Third, the estimation of a mixture or a distribution of the particles with different sizes can be important for some scenarios, such as aggregation of probes or during polymer degradation processes. \newedit{As the particles with different sizes may have different thermal fluctuations, the correlation patterns modeled in ISF for these particles is also different. It is of interest to include a mixture or a distribution of ISFs in the AIUQ framework for systems with inhomogeneous particle sizes. \revise{Lastly, certain particles can undergo rapid photobleaching in some experiments, which can alter the signal-to-noise ratio. Thus, it is of interest to model time-dependent amplitude,  $A(\mathbf q, t)$, and noise parameters $\bar B(t)$ for these scenarios. 
}}}

\quad 

\section{Acknowledgements}
This project is partially supported by the National Science Foundation under Grant No. OAC-2411043. X.L. is supported by the National Science Foundation under Grant No.
DMS-2053423.  M.G. acknowledges the supplement to National Science Foundation  Grant No. 2119663. 
Y.H. acknowledges partial support from the UC  Multicampus Research Programs and Initiatives (MRPI) program under Grant No. M23PL5990  and from the BioPACIFIC Materials Innovation Platform of the National Science Foundation under Grant No. DMR-1933487 (NSF BioPACIFIC MIP). 
Y.L. acknowledges the support from the National Science Foundation under Grant No. OAC-2411044 and the donors of ACS Petroleum Research Fund under Doctoral New Investigator No. 67734-DNI10. The authors thank Matthew Helgeson, Ryan McGorty,  Rae Robertson-Anderson and Megan Valentine for helpful discussions. 

\bibliography{References_chronical_2022}
\section*{Appendices}
\setcounter{equation}{0}
\renewcommand{\theequation}
{A\arabic{equation}}

\section*{Appendix A: Derivation for Section \ref{sec:background}}

The intermediate scattering function (ISF) defined in Eq. (\ref{equ:ISF}) is computed by
\begin{align}
&f_{\bm \theta}(\mathbf q,\Delta t) \nonumber\\
=&\mbox{Cov}(\psi(\mathbf q,t), \psi^*(\mathbf q,t+\Delta t)) \nonumber\\
=&\mathbb E[\psi(\mathbf q,t)\cdot\psi^*(\mathbf q,t+\Delta t)] \nonumber \\
=& \mathbb E\left[\frac{1}{M}\sum^M_{m=1} \exp\left(i\mathbf q \cdot (\mathbf x_m(t+\Delta t)-\mathbf x_m(t)) \right) \right] + \nonumber \\
& \qquad \mathbb E\left[\frac{1}{M}\sum_{m\neq m'} \exp\left(i\mathbf q \cdot (\mathbf x_m(t+\Delta t)-\mathbf x_{m'}(t)) \right) \right] \nonumber \\
=&\Bigl< \frac{1}{M}\sum^M_{m=1} \exp\left( i\mathbf q \cdot \Delta \mathbf  x_m(t,\Delta t)\right) \Bigr>. \nonumber
\end{align}     
The last equation holds because there is no interaction between particles and the long-term expectation of $\psi$ is zero by assumption \cite{berne2000dynamic}. Note that the derivation of ISF assumes the particle intensity profile is a Dirac delta function, which does not strictly hold in light microscopy. 
Consequently, the amplitude is relevant to various factors, such as the structure factor and the particle form factor \cite{nixon2022probing} and they typically need to be estimated.

As all particles have the same displacement distributions, the ISF of 2D processes can be written as
\begin{align}
f_{\bm \theta}(\mathbf q, \Delta t)\nonumber & =\mathbb E\left[ \exp\left( i\mathbf q \cdot \Delta \mathbf x(t,\Delta t)\right)\right] \nonumber \\
&=\mathbb E\left[ \exp\left( i q_1  \Delta  x_1(t,\Delta t)\right)\right]\mathbb E\left[ \exp\left( i q_2  \Delta  x_2(t,\Delta t)\right)\right] \nonumber \\
&:=f_{\bm \theta_1}(q_1,\Delta t) f_{\bm \theta_2}(q_2,\Delta t).
\label{equ:f_f1_f2}
\end{align}
Based on the cumulant generating function, for any $l=1,2$, we have a power series expansion of the ISF  \cite{nijboer1966time,koppel1972analysis},  
\begin{align}
\log(f_{\bm \theta_l}(q_l,\Delta t))=\sum^{\infty}_{j=1} \kappa_{l,j}\frac{ (iq_l)^j  }{j!},
\label{equ:cumulant_ISF}
\end{align}
where $\kappa_{l,j}=f^{(j)}_{\bm \theta_l}(q_l, 0)$ is the $j$th derivative of the ISF with respect to the random displacement along the $l$th coordinate at time $\Delta t$. 
Note that $\kappa_{l,1}=\mathbb  E[\Delta  x_l(t,\Delta t)]$ and  $\kappa_{l,2}=\mathbb E[\Delta x_l(t, \Delta t)^2]$ are first and second moments, respectively. As the first moment is zero,  approximating the log ISF along the $l$th coordinate by the  first two moments follows
\begin{equation}
\log(f_{\bm \theta_l}(q_l,\Delta t))\approx \frac{ -q^2_l \mathbb E[\Delta x^2_l(t,\Delta t)] }{2}. 
\label{equ:f_l}
\end{equation}
\edit{When random displacements are Gaussian, there is no approximation. In general,  approximation in Eq. (\ref{equ:f_l}) is widely used in related techniques such as dynamic light scattering \cite{stetefeld2016dynamic}. Noting that we only use MSD-parameterized ISF as examples. The AIUQ approach is applicable to all ISFs that may not be approximated by MSD.  }

Substitute $f_l(q_l,\Delta t)$ from Eq. (\ref{equ:f_l}) into (\ref{equ:f_f1_f2}) gives an approximation of the ISF by the MSD in a 2D space:  
\begin{align}
&f_{\bm \theta}(\mathbf q, \Delta t) \nonumber \\
\approx &\exp\left\{-\frac{1}{2} \mathbb E\left[q^2_1  \Delta x^2_1(t,\Delta t)+q^2_2  \Delta x^2_2(t, \Delta t) \right]\right\} 
\label{equ:f_theta_anisotropic}\\
= & \exp \left\{ -\frac{1}{4}\mathbb E\left[(q^2_1+q^2_2)(\Delta x^2_1 (t,\Delta t)+\Delta x^2_2(t,\Delta t)) \right] \right\} \nonumber \\
=& \exp\left(-\frac{q^2 \langle \Delta   x^2(\Delta t)\rangle}{4} \right), \nonumber 
\end{align}
where $q^2=q^2_1+q^2_2$ and the MSD $\langle\Delta  x^2(\Delta t)\rangle=\mathbb E\left[\Delta x^2_1 (t,\Delta t)+\Delta x^2_2(t,\Delta t)\right]$. 

 \begin{table}[t]
\centering
\begin{tabular}{ll}
\hline
Parametric  & ISF \\ 
\hline
BM  &  $\exp\left(- q^2 \sigma^2_{BM} \Delta t/4 \right)$ \\ 
FBM  &  $\exp\left(-q^2 \sigma^2_{FBM} \Delta t^{\alpha}/4 \right)$  \\ 
 OU &$\exp\left(-q^2 \sigma^2_{OU}(1-\rho^{\Delta t})/4 \right)$ \\ 
 OU$+$FBM  &$\exp\left(-q^2 (\sigma^2_{1} \Delta t^{\alpha}+\sigma^2_{2}(1-\rho^{\Delta t}) )/4 \right)$ \\ 
 \hline
 Nonparametric  & ISF \\ 
 \hline
Cumulant approx. & $\exp\left(-q^2  \langle \Delta x^2(\Delta t) \rangle/4 \right)$ \\
\hline
\end{tabular}
\caption{A list of parametric models of the intermediate scattering function (ISF) for Brownian motion (BM), fractional Brownian motion (FBM), Ornstein–Uhlenbeck (OU) process, and a mixture of the OU process and FBM (OU$+$FBM). 
The nonparametric model uses cumulant approximation \cite{koppel1972analysis} to construct the ISF by the mean squared displacement $ \langle \Delta x^2(\Delta t) \rangle$, which gives a unique parameter at any lag time $\Delta t$. 
  }
\label{tab:ISF_table}
\end{table}

ISFs of a few widely used processes are summarized in Table \ref{tab:ISF_table}, which follows from approximation by MSD.   For a Brownian motion (BM), for instance, the update of the 2D position of the $m$th particles follows $\mathbf x_{m}(t+\Delta t_{min})=\mathbf x_{m}(t)+\frac{\sigma^2_{BM}}{2} \bm \epsilon_{m}(t)$ with $\bm \epsilon_{m}(t)\sim \mathcal{MN}(\mathbf 0,\mathbf I_2)$.
 The MSD of BM follows $\mbox{MSD}_{BM}=\sigma^2_{BM}\Delta t$ \cite{gu2021uncertainty}. 
 
 For an OU process, the particle's successive steps have a weaker correlation with previous steps than the BM, 
 \begin{align}
   \mathbf x_m(t+\Delta t_{min})&=\rho(\mathbf  x_m(t)- \mathbf  x_m(t_1))+\mathbf  x_m(t_1) \nonumber\\
   &\quad \quad +\frac{\sigma_{OU}^2(1-\rho^2)}{4}\bm \epsilon_m(t),
   \label{eq:sim_OU}
 \end{align}
where $\mathbf x_m(t_1) \sim \mathcal{MN}(\mathbf x_m(t_0), \frac{\sigma_{OU}^2}{4} \mathbf I_2)$, for a deterministic position $\mathbf x_m(t_0)$. The  MSD of the OU process follows  $\sigma_{OU}^2(1-\rho^{\Delta t})$ \cite{gu2021uncertainty}. 

We introduce the MSD and derivation for two other processes: 
the fractional Brownian motion (FBM), 
and a mixture of the OU process and FBM, which has the same MSD as the continuous time random walk (CTRW) and the noisy continuous time random walk (NCTRW), respectively \cite{jeon2013noisy,metzler2014anomalous}. Denote $\Delta n = \Delta t/\Delta t_{min}$. First, simulated particles from a 2D FBM process exhibit long-term dependence, and the self-similarity is controlled by the Hurst parameter $H=\alpha/2$ \cite{mandelbrot1968fractional,decreusefond1999stochastic}:
\begin{equation}
    x_{m,l}(t+\Delta t) = x_{m,l}(t)+ 
    \tilde x_{m,l}(\Delta t), 
    \label{eq:fbm_sim}
\end{equation}
where the incremental process  
$\tilde x_{m,l}(\Delta t) $ is known as the fractional Gaussian noise with $\mathbb E\left[\tilde x_{m,l}(\Delta t)\right] = 0$, and $\Cov\left(\tilde x_{m,l}(\Delta t_k),\tilde x_{m,l}(\Delta t_s)\right)=\frac{\sigma_{FBM}^2}{4}\left(|\Delta t_k|^{2H}+|\Delta t_s|^{2H}-|\Delta t_k-\Delta t_s|^{2H}\right)$; the indices representing the $m$th particle and $l$th direction ($l=1,2$) and $\Delta t_k=k\Delta t_{min}$ and $\Delta t_s=s\Delta t_{min}$  for any integer $s$ and $k$.  
The MSD in one coordinate can be computed as 
\begin{align*}
    &\mathbb E\left[\left(x_{m,l}(t+\Delta t)-x_{m,l}(t)\right)^2\right]\\
    =&\mathbb V\left[x_{m,l}(t+\Delta t)-x_{m,l}(t)\right] +\mathbb E^2\left[x_{m,l}(t+\Delta t)-x_{m,l}(t)\right]\\
    =&\mathbb V\left[\tilde x_{m,l}(\Delta t)\right]=\frac{\sigma_{FBM}^2}{2}\Delta t^{2H}=\frac{\sigma_{FBM}^2}{2}\Delta t^{\alpha}, 
\end{align*} 
 for any  particle $m = 1,\dots,M$ and coordinate $l=1,2$. 
Since particles move isotropically in a 2D space, the MSD for FBM is  $\mathbb E\left[ \Delta x^2(\Delta t) \right ] = \sigma_{FBM}^2\Delta t^{\alpha}$. 

We next consider particles undergoing a mixture of the OU process and FBM. This model presents a generalized form of FBM and it has the same form of MSD as the NCTRW \cite{metzler2014anomalous}.
This extension is suitable for describing scenarios where the particles are confined within cages and exhibit waiting times distributed according to power laws.  Here, the process can be simulated through a summation of a mutually independent  FBM and an OU process. The update of the particles' position follows  
\begin{equation}
      x_{m,l}(t) =  u_{m,l}(t)+v_{m,l}(t), 
    \label{eq:ou-fbm}
\end{equation}
 where $u_{m,l}(t)$ is a FBM process in the $l$th direction with MSD $\sigma^2_1\Delta t^{\alpha}/2$,  
 and $v_{i,l}(t)$ stands for an independent OU process in the $l$th direction with MSD $\sigma^2_2(1-\rho^{\Delta t})/2$.   
Since particles move isotropically in a 2D space, and the two processes are independent, the MSD is $\sigma^2_{1}\Delta t^{\alpha}+\sigma_{2}^2(1-\rho^{\Delta t})$.


\section*{Appendix B: q-dependent estimation when directly  inverting the image structure function}



\edit{We show another example to indicate the estimation in model-free analysis of DDM depends on the choice of wave-vector range \cite{bayles2017probe,edera2017differential,gu2021uncertainty}  using a diffusive process or Brownian motion. Here the MSD $\langle \Delta x^2(\Delta t) \rangle$ is obtained by directly inverting the observed image structure function $D(q,\Delta t)$:
\begin{align}
  \Delta x^2_{est}(q, \Delta t) &= \frac{4}{q^2}\mbox{log} \left[ \frac{A(q)}{A(q)-D(q,\Delta t) + B(q,\Delta t)} \right]. \nonumber \\
\label{equ:msd}
\end{align}}

\begin{figure} 
\centering
\includegraphics[width=0.48\textwidth]{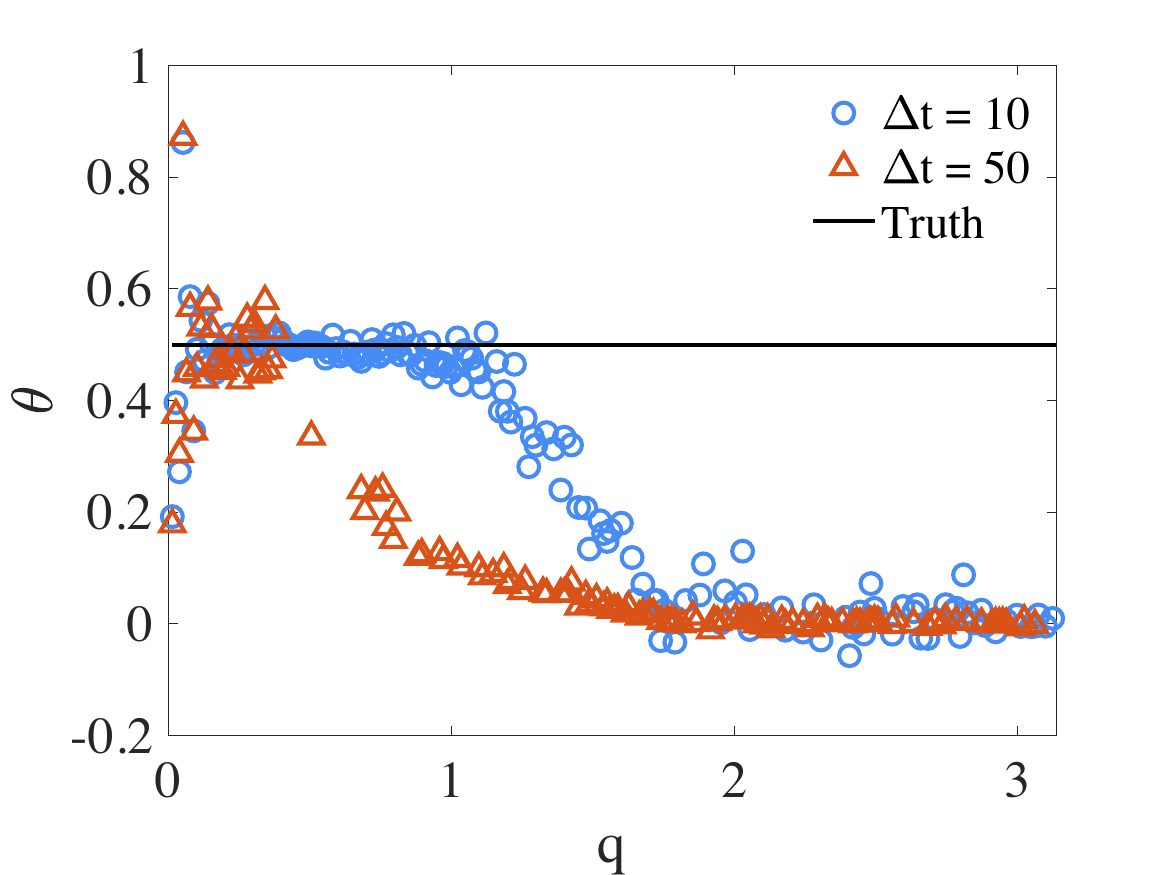}
   \caption{Parameter estimation by directly inverting $D(q,\Delta t)$ using Eq. \ref{equ:msd}.  Blue circles and red triangles denote estimation of the parameter at $\Delta t$ = 10 and 50, respectively. The truth ($\theta = 0.5$) is denoted by the thick black line. }
   \label{fig:direct_inverstion_dqt}
\end{figure}

\edit{We simulate the diffusive process with  $500\times 500$ pixels across $ 500$ frame with diffusion coefficient $\theta=0.5$ in Figure \ref{fig:direct_inverstion_dqt}. 
We follow the first approach in fitting $D(q,\Delta t)$ to estimate $A_j$ and $\bar B$ discussed in Sec \ref{subsec:comparison_mmle_fitting}. 
For purely diffusive motion, $\langle \Delta x^2(\Delta t) \rangle = 4\theta \Delta t$,  containing only one parameter with the truth $\theta = 0.5$. 
At $\Delta t = 10$, roughly one-third of $q$ are close to the estimate of $\theta$, while at $\Delta t = 50$, a much narrower range  $q$ produces values close to the true diffusion coefficient. This highlights the need for selecting a wave-vector range and weighing the contribution of different $q$'s even for diffusive processes.
}

\section*{Appendix C: Derivation for Section \ref{sec:latent_factor}}
We first derive the distribution of $\Delta {\hat y}_{re,\mathbf q}(t,\Delta t)$ \revise{for any $\mathbf q$, t, and $\Delta t$}. The distribution of $\Delta {\hat y}_{im,\mathbf q}(t,\Delta t)$ can be shown similarly. 
Note $\mathbb E[\Delta {\hat y}_{re,\mathbf q}(t,\Delta t)]=0$. The variance can be computed by 
\begin{align*}
&\mathbb V[\Delta {\hat y}_{re,\mathbf q}(t,\Delta t)]\\
=&\mathbb V[{\hat y}_{re,\mathbf q}(t+\Delta t)]+\mathbb V[{\hat y}_{re,\mathbf q}(t)] \\
&\qquad-2 \mbox{Cov}[{\hat y}_{re,\mathbf q}(t), {\hat y}_{re,\mathbf q}(t+\Delta t)] \\
=&2\times\frac{A(\mathbf q)}{4} +2\times \frac{\bar B}{4} -2 \times\frac{A(\mathbf q)}{4}f_{\bm \theta}(\mathbf q,\Delta t)\\
=&\frac{A(\mathbf q)}{2}(1-f_{\bm \theta}(\mathbf q,\Delta  t)) +\frac{\bar B}{2}.
\end{align*}
As $\Delta {\hat y}_{re,\mathbf q}(t,\Delta t)$ follows a Gaussian distribution, we have $\Delta {\hat y}_{re,\mathbf q}(t,\Delta t)\sim\mathcal N\left(0, \frac{A(\mathbf q)}{2}\left(1-f_{\bm \theta}(\mathbf q,\Delta  t)\right) +\frac{\bar B}{2} \right)$. 

We next derive the probability density of $J$ rings of transformed intensity at the Fourier space in Eq. (\ref{equ:prod_density}). 
From Eq. (\ref{equ:y_re_hat}), the mean of an $n$-vector $\hat{\mathbf y}_{re,\mathbf j'}=(\hat{y}_{re,\mathbf j'}(t_1),...,\hat{y}_{re,\mathbf j'}(t_n))^T$ for any $\mathbf j'\in \mathcal S_j$ can be computed by the law of total expectation below: 
\begin{align}
\mathbb E[\mathbf {\hat y}_{re,\mathbf j'}] =\mathbb E[\mathbb E[\mathbf {\hat y}_{re,\mathbf j'}\mid \mathbf z_{re,\mathbf j'}]]
 =\mathbb E[\mathbf z_{re,\mathbf j'}]=\mathbf 0,
 \label{equ:mean_hat_y_re_j}
\end{align}
where $\mathbf z_{re,\mathbf j'}=(z_{re,\mathbf j'}(t_1),...,z_{re,\mathbf j'}(t_n))^T$ is a vector of zero-mean, real-valued random factors. The covariance of $\hat{\mathbf y}_{re,\mathbf j'}$  can be computed by the law of total covariance: 
\begin{align}
\mathbb V[\mathbf {\hat y}_{re,\mathbf j'}]&=\mathbb V[\mathbb E[\mathbf {\hat y}_{re,\mathbf j'}\mid \mathbf z_{re,\mathbf j'}]]+\mathbb E[\mathbb V[\mathbf {\hat y}_{re,\mathbf j'}\mid \mathbf z_{re,\mathbf j'}]]\nonumber \\
&=\mathbb V[\mathbf z_{re,\mathbf j'}]+\frac{\bar B}{4}\mathbf I_n
=\frac{A_j}{4}\mathbf R_j+\frac{\bar B}{4}\mathbf I_n.
 \label{equ:var_hat_y_re_j}
\end{align}
Note that $\mathbf {\hat y}_{re,\mathbf j'}$ is Gaussian since both $\mathbf z_{re,\mathbf j'}$ and $\bm \epsilon_{j,re}=(\bm \epsilon_{j,re}(t_1),...,\bm \epsilon_{j,re}(t_n))^T$ are Gaussian. Furthermore, the random factors $\mathbf z_{re,j'_1}$ and $\mathbf z_{re,j'_2}$ are independent when $j'_1\neq j'_2$, and the noise is also independent. Hence the probability density of $\tilde N\times n$ matrix $\mathbf{\hat y}_{re}=[\mathbf {\hat y}_{re}(t_1),...,\mathbf {\hat y}_{re}(t_n)]$ follows 
\begin{equation*}
p(\mathbf {\hat y}_{re}\mid \bm \theta, \mathbf A_{1:J},\bar B)=\prod^{J}_{j=1} \prod_{\mathbf j' \in \mathcal S_j} p_{MN}\left(\mathbf {\hat y}_{re,\mathbf j'};\, \mathbf 0,\, \bm \Sigma_j \right). 
\end{equation*}
As the density of the imaginary part can be similarly derived, we have the logarithm of the likelihood below  
 \begin{align}
 &\log(\mathcal L\left(\bm \theta,\bm A_{1:J}, \bar B \right)) \nonumber \\
 = & -{n\tilde N}\mbox{log}(2\pi) -\sum^{J}_{j=1} \left\{{S_j} \log(|\bm \Sigma_j|)  \right\} \nonumber \\
 &  - \frac{S_j}{2}\sum_{\mathbf j' \in \mathcal S_j}  \big(\mathbf {\hat y}_{re,\mathbf j'}^T \bm \Sigma_j^{-1}\mathbf {\hat y}_{re,\mathbf j'}  + \mathbf {\hat y}_{im,\mathbf j'}^T \bm \Sigma_j^{-1}\mathbf {\hat y}_{im,\mathbf j'} \big),
 \label{equ:log_density}
 \end{align} 
 which is the density of the right-hand-side of Eq. (\ref{equ:prod_density}). 

 \revise{Lastly, we provide the derivation to prove  the unbiasedness of the estimator $A_{est,j}$ in Eq. (\ref{equ:A_est_j}). Note that for any $\mathbf j'\in \mathcal S_j$ the mean and covariance of $\mathbf {\hat y}_{re,\mathbf j'}=({\hat y}_{re,\mathbf j'}(t_1),...,{\hat y}_{re,\mathbf j'}(t_n))^T$ and  $\mathbf {\hat y}_{im,\mathbf j'}=({\hat y}_{im,\mathbf j'}(t_1),...,{\hat y}_{im,\mathbf j'}(t_n))^T$ are the same.  
 Then, by Eq. (\ref{equ:mean_hat_y_re_j}) and Eq. (\ref{equ:var_hat_y_re_j}), for any  $\mathbf j' \in \mathcal S_j$ and $t_k$ with $1\leq k\leq n$ we have 
 \begin{align}
 \mathbb E[ |\hat {y}_{\mathbf j'}(t_k)|^2 ]&=  \mathbb E[ \hat { y}_{re,\mathbf j'}(t_k)^2+\hat { y}_{im,\mathbf j'}(t_k)^2  ]\nonumber \\
 &=2\times \left(\frac{A_j}{4}+\frac{\bar B}{4}\right)\nonumber \\
 &=\frac{A_j+\bar B}{2}. 
 \end{align}
Then  
\begin{align}
\mathbb E[A_{est,j}]&=\frac{2}{ S_j n} \sum_{\mathbf j'\in \mathcal S_j}\sum^{n}_{k=1}\mathbb E[|\hat y_{\mathbf j'}(t_k)|^2]-{\bar B}\nonumber\\
&=2\times \left(\frac{A_j+\bar B}{2}\right)-\bar B=A_j. 
\end{align}
Since the expected value of the estimator $A_{est,j}$ is equal to the underlying true value, the estimator $A_{est,j}$ is unbiased. 
 }

\end{document}